\documentclass{emulateapj}

\bibpunct{(}{)}{;}{a}{}{,}

\begin{document}

\title{X-ray emission from Planetary Nebulae \\ 
I. Spherically symmetric numerical simulations}

\author{Matthias Stute and Raghvendra Sahai}
\affil{Jet Propulsion Laboratory, California Institute of Technology, 
4800 Oak Grove Drive, Pasadena, CA 91109, USA}
\email{Matthias.Stute@jpl.nasa.gov, \\ Raghvendra.Sahai@jpl.nasa.gov}

\begin{abstract}
The interaction of a fast wind with a spherical Asymptotic Giant Branch 
(AGB) wind is thought to be the basic mechanism for shaping 
Pre-Planetary Nebulae (PPN) and later Planetary Nebulae (PN). Due to the large 
speed of the fast wind, one expects extended X-ray emission from these objects,
but X-ray emission has only been detected in a small fraction of PNs and only 
in one PPN.
Using numerical simulations we investigate the constraints that can be set on 
the physical properties of the fast wind (speed, mass-flux, opening 
angle) in order to produce the observed X-ray emission properties of PPNs and 
PNs. 
We combine numerical hydrodynamical simulations including radiative cooling 
using the code {\em FLASH} with calculations of the X-ray properties of the 
resulting expanding hot bubble using the atomic database {\em ATOMDB}.
In this first study, we compute X-ray fluxes and spectra using 
one-dimensional models. Comparing our results with analytical solutions,
we find some 
agreements and many disagreements. In particular, we test the effect of 
different time histories of the fast wind on the X-ray emission and find that 
it is determined by the final stage of the time history during which the fast 
wind velocity has its largest value. The disagreements which are 
both qualitative and quantitative in nature argue for the necessity of using 
numerical simulations for understanding the X-ray properties of PNs. 
The X-ray luminosity in the 0.2 -- 10 keV 
range (1.24 -- 62 \AA) covered by the CHANDRA/{\em ACIS} instrument shows a 
very slow decrease 
with time over the range of evolutionary ages explored in our models (up to 
3750 years).
Furthermore, investigating the emission in other wavelengths ranges, we find 
that most of the luminosity emerges at longer wavelengths 
($\lambda \gtrsim 140$ \AA) from the cooler outer edge of the hot bubble. 
We apply our spherical models to the objects BD$+$30$^\circ$3639 and 
NGC~40. We find that the model values of the X-ray temperature and 
luminosity for these objects are significantly higher than observed values 
and discuss several mechanisms for resolving the discrepancies.
\end{abstract}

\keywords{circumstellar matter --- hydrodynamics --- ISM: jets and outflows ---
planetary nebulae: general --- stars: mass loss --- X-rays: ISM}

\section{Introduction}

The shaping of Pre-planetary nebulae (PPN) and Planetary Nebulae (PN) is 
believed to result from the interaction of a fast, collimated post-AGB wind 
plowing into the slow, dense wind emitted during the AGB phase, followed 
by an isotropic tenuous wind during the PN phase \citep{SaT98}. These 
interacting winds result in an expanding shell of shocked matter which forms 
the PPN and later the PN. Due to the large speed of the fast wind (from few x 
100 km s$^{-1}$ up to 2000 km s$^{-1}$), one expects extended X-ray emission 
to be produced in PPNs (where the shaping mainly occurs) and PNs. 

However, X-ray emission was only detected in 3 of 60 PNs observed with ROSAT, 
followed by a few more from CHANDRA and XMM \citep[e.g.][]{GCG05}. In the case 
of PPNs, there is so far only one confirmed X-ray detection
\citep{SKF03}, although many have been observed with CHANDRA. X-ray emission 
should be present in principle in all of these systems -- hence the low 
detection rate implies that it is obviously below the detection limit. 

The general problem of understanding the formation and shaping of PNs has been 
addressed analytically and numerically 
\citep[see][and references therein]{BaF02}. 
In principle, one of the most direct probes of the fast wind and the 
interaction process which drives PN formation is the X-ray emission. 
It may help in answering several important questions -- e.g. what are the 
physical properties of the fast wind (speed, mass-flux, opening angle -- and 
the time histories of these parameters). Such studies can also help in 
identifying the nebular regions that are mainly responsible for the X-ray 
emission for comparison with recent X-ray images of PNs obtained with CHANDRA 
and XMM \citep[e.g.][]{KSV00,KBB03,GCG05}.

X-ray emission as a probe was first exploited by \citet{ZhP96}, who derived 
an analytical, self-similar, spherically symmetric model, taking into account 
the effects of thermal conductivity. Very recently, 
\citet[][hereafter ASB06]{ASB06} presented new analytical models with 
radiative cooling, but without heat conduction. Although analytical modeling 
is potentially a powerful tool for understanding of the X-ray emission from PNs
and its dependence on fundamental physical parameters, it necessitates making 
certain assumptions which may or may not be completely valid. For example, a 
major problem with such analytical models is, that they are not able to treat 
all the effects (hydrodynamics, radiative cooling, heat conduction) 
self-consistently. In this paper, we have therefore performed numerical 
simulations using the code {\em FLASH} in which we vary the basic parameters 
of the fast and slow wind over an extensive parameter grid and compute the 
X-ray emission as a function of these parameters. A major goal of our work here
is to test the main results from ASB06's analytical modeling and verify their  
general conclusions. We have therefore, as in ASB06, not included heat 
conduction in our treatment. However, we discuss the heat conduction process 
and how it may be inhibited by magnetic fields which are likely to be present 
in PNs. For example, the lack of spherical symmetry in most PNs has been
explained by some authors as resulting from the presence of magnetic fields
\citep[e.g.][]{GLF05}. This paper, in which we have carried out spherically 
symmetric simulations of interacting winds, is a first in a series of papers 
in our quest to understand X-ray emission from PPNs and PNs. 

The rest of the paper is organized as follows. In \S \ref{sec_model}, we 
describe our numerical model. In \S \ref{sec_res_1D}, we present the results 
of our one-dimensional simulations including the structure and X-ray properties
of the expanding bubble and their dependency on the parameters of the fast 
wind. In \S \ref{sec_res_2D}, new effects from the computation of a 
spherically symmetrical model on a two-dimensional grid are shown. In \S 
\ref{sec_appl}, we apply our models to two PNs, BD$+$30$^\circ$3639 and NGC~40,
and examine several mechanisms for resolving the discrepancies between the 
models and observed values of the X-ray temperature $T_{\rm x}$ and 
luminosity $L_{\rm x}$ for these objects.
We compare in detail our results with those of ASB06 in \S \ref{sec_diff} and
discuss the still unsolved problem of fitting $T_{\rm x}$ and $L_{\rm x}$ in 
PNs in \S \ref{sec_fit}. Finally in \S \ref{sec_concl}, we present our 
conclusions.

\section{The numerical method} \label{sec_model}

\subsection{Initial and boundary conditions}

Using {\em FLASH}, we solve the following set of the differential equations of 
ideal hydrodynamics 
\begin{eqnarray} \label{hydro}
\frac{\partial\,\rho}{\partial\,t} + \nabla\,(\rho\,{\bf v}) &=& 0 \nonumber \\
\frac{\partial\,(\rho\,{\bf v})}{\partial\,t} + \nabla\,
(\rho\,{\bf v}\otimes{\bf v}) &=& - \nabla\,p \nonumber \\
\frac{\partial\,e}{\partial\,t} + \nabla\,(e\,{\bf v}) &=& - p\,\nabla\,{\bf v}
 - n^2\,\Lambda ( T ) \nonumber \\
p &=& (\gamma - 1)\,e .
\end{eqnarray}
\noindent
$\rho$ is the gas density, $p$ the pressure, $e$ the internal energy density, 
${\bf v}$ the velocity and $\gamma$ the ratio of the specific heats at 
constant pressure and volume which is set to $\gamma = 5/3$. We include 
radiative losses using the cooling function $\Lambda ( T )$ of \citet{SuD93} 
in the temperature range between $10^4$ and $10^{8.5}$ K and the low 
temperature function of \citet{DaM72} for $T<10^4$ K.

A spherically symmetric AGB wind is implemented with
\begin{equation}
\rho_{\rm s} = \frac{\dot M_{\rm s}}{4\,\pi\,r^2\,v_{\rm s}} \equiv \rho_{01}\,
r^{-2}
\end{equation}
and 
\begin{equation}
\vec v_{\rm s} = v_{\rm s} \, \vec e_{r} .
\end{equation}
The temperature of the AGB wind is set to 10 K. In this $r^{-2}$-density 
profile, we let a collimated fast wind (CFW) plow from the inner radial 
boundary with the following parametrization:
\begin{equation} \label{eq_rho_CFW}
\rho_{\rm f} = \frac{\dot M_{\rm f}}{4\,\pi\,r^2\,v_{\rm f}\,K}\,
\exp \left( - \left[ \frac{\theta}{\theta_{\rm f}} \right]^2 \right)
\end{equation}
and 
\begin{equation} \label{eq_v_CFW}
\vec v_{\rm f} = v_{\rm f}\,\exp \left( - \left[ \frac{\theta}{\theta_{\rm f}} 
\right]^2 \right)\,\vec e_{r} .
\end{equation}
$\theta_{\rm f}$ is the opening angle of the CFW. K is a normalization factor 
calculated by
\begin{equation}
K = \int_{0}^{\pi/2}\,\exp \left( - 2\,\left[ \frac{\theta}{\theta_{\rm f}} 
\right]^2 \right)\,\sin\,\theta\,d\theta .
\end{equation}
For a temperature of the CFW, we assume $10^4$ K. The simulations are 
performed in spherical coordinates with the dimensions of 
$r = 5\times10^{13}$ -- $2.5 \times 10^{17}$ cm and 
$\theta = 0$ -- $\pi/2$ in the case of the two-dimensional runs. Mirror 
boundaries are used along the symmetry axis and the equatorial plane and 
outflow conditions at the upper radial boundary.

\subsection{Model parameters} \label{sec_param}

The models which are presented in this paper are spherical 
ones without the $\theta$ dependency in eqns. \ref{eq_rho_CFW} -- 
\ref{eq_v_CFW} and $K = 1$. Preliminary results of models with $\theta$ 
dependency (i.e. incorporating collimated fast winds) were published in 
\citep{StS06}. Detailed results are deferred to 
a future publication. The parameters of the models are then the mass 
outflow rate and velocity of the fast and the slow wind, respectively. For 
these parameters we adopted the same ranges as in the models of ASB06 for ease 
of comparison (see Table 1). Thus we performed three sets of runs with 
$\dot M_{\rm s}$ of $3\times10^{-6}$ (runs A), $7\times10^{-6}$ (runs B) and 
$3\times10^{-5}$ M$_\odot$ yr$^{-1}$ (runs C) and within each set we varied 
the velocity of the fast wind from 300 to 700 km s$^{-1}$ (e.g. the 5 in C5 
means $v_{\rm f}=500$ km s$^{-1}$). The velocity of the slow wind was set to 
10 km s$^{-1}$ and $\dot M_{\rm f}$ was chosen such that 
$\dot M_{\rm f}\,v_{\rm f} = \dot M_{\rm s}\,v_{\rm s}$. We expanded these
sets while applying our models to observations.
\begin{deluxetable*}{ccccccccccc}
\tablehead{\colhead{run} & \colhead{$\dot M_{\rm f}$} & 
\colhead{$v_{\rm f}$} & \colhead{$\dot M_{\rm s}$} & 
\colhead{$v_{\rm s}$} & \colhead{$v_{\rm cr}$} & 
\colhead{$R_{\rm cd}^{\rm E} / t$} & \colhead{$R_{\rm cd}^{\rm M} / t$} & 
\colhead{$\mathcal{I}$} & \colhead{$v_{\rm ASB}$} & 
\colhead{$v_{\rm cd, sim}$}
}
\startdata
A3 & $1\times10^{-7}$ & 300 & $3\times10^{-6}$ & 10 & 21.54 & 27.51 & 
9.58 & 0.033 & 28.6 & 25.16 \\
A5 & $6\times10^{-8}$ & 500 & $3\times10^{-6}$ & 10 & 25.54 & 32.62 & 
9.8 & 0.02 & 32.6 & 28.47 \\
A7 & $4.33\times10^{-8}$ & 700 & $3\times10^{-6}$ & 10 & 28.67 & 36.62 & 
9.66 & 0.014 & 35.6 & 30.92 \\
\tableline
B3 & $2.33\times10^{-7}$ & 300 & $7\times10^{-6}$ & 10 & 21.53 & 27.51 & 
9.58 & 0.033 & 28.6 & 23.97 \\
B4 & $1.75\times10^{-7}$ & 400 & $7\times10^{-6}$ & 10 & 23.71 & 30.28 & 
9.75 & 0.025 & 30.8 & 25.70 \\
B5 & $1.4\times10^{-7}$ & 500 & $7\times10^{-6}$ & 10 & 25.54 & 32.62 & 
9.8 & 0.02 & 32.6 & 26.96 \\
B6 & $1.17\times10^{-7}$ & 600 & $7\times10^{-6}$ & 10 & 27.17 & 34.7 & 
10.03 & 0.017 & 34.2 & 28.10 \\
B7 & $1\times10^{-7}$ & 700 & $7\times10^{-6}$ & 10 & 28.58 & 36.49 & 
9.66 & 0.014 & 35.6 & 29.10 \\
\tableline
C3 & $1\times10^{-6}$ & 300 & $3\times10^{-5}$ & 10 & 21.54 & 27.51 & 
9.58 & 0.033 & 28.6 & 22.60 \\
C5 & $6\times10^{-7}$ & 500 & $3\times10^{-5}$ & 10 & 25.54 & 32.62 & 
9.8 & 0.02 & 32.6 & 25.05 \\
C7 & $4.28\times10^{-7}$ & 700 & $3\times10^{-5}$ & 10 & 28.57 & 36.47 & 
9.66 & 0.014 & 35.6 & 26.88 \\
\tableline
A10 & $3\times10^{-8}$ & 1000 & $3\times10^{-6}$ & 10 & 32.18 & 41.10 & 9.90 
& 0.01 & -- & 36.56 \\
B10 & $7\times10^{-8}$ & 1000 & $7\times10^{-6}$ & 10 & 32.18 & 41.10 & 9.90 
& 0.01 & -- & 33.43 \\
D7 & $1\times10^{-6}$ & 700 & $7\times10^{-5}$ & 10 & 28.58 & 36.49 & 9.66 & 
0.014 & -- & 25.99 \\
E10 & $2.4\times10^{-6}$ & 1000 & $5\times10^{-4}$ & 10 & 25.19 & 32.17 & 6.88
& 0.007 & -- & 27.03\\
\tableline
B5 (2D) & $1.4\times10^{-7}$ & 500 & $7\times10^{-6}$ & 10 & 25.54 & 32.62 & 
9.8 & 0.02 & -- & 22.4 \\
\enddata
\tablecomments{The first five columns give the parameters of our models: the 
name of the run, the mass outflow rate and velocity of the fast and the slow 
wind, respectively. The next four columns give results of the analytical 
solution of \citet{KMK92a,KMK92b}: the critical velocity, the velocity of 
the contact discontinuity in the energy-driven case and in the momentum-driven 
case and the constant $\mathcal{I}$. The velocity used by ASB06 following a 
formula in \citet{VoK85} is given in the column before last, the last column 
shows the measured velocity of the contact discontinuity $v_{\rm cd, sim}$ in 
the simulations.}
\end{deluxetable*}

\subsection{Calculating the X-ray properties} \label{sec_calc_xray}

Using the radial density and temperature profiles from the hydrodynamical 
simulations, we determine the expected X-ray flux. We have used the atomic 
database {\em ATOMDB} with {\em IDL} including the Astrophysical Plasma 
Emission Database ({\em APED}) and the spectral models output from the 
Astrophysical Plasma Emission Code ({\em APEC}) \citep{SBL01} to calculate the 
emissivity. The default abundances in {\em ATOMDB}, i.e. 14 elements (H, He, C,
N, O, Ne, Mg, Al, Si, S, Ar, Ca, Fe, Ni) with solar abundances of 
\citet{AnG89} are used. The energy range is divided into bins of 0.01 keV.
We compute the spectrum and the total flux in X-rays as a function of 
evolutionary time for each of our models.

\begin{figure}
\plotone{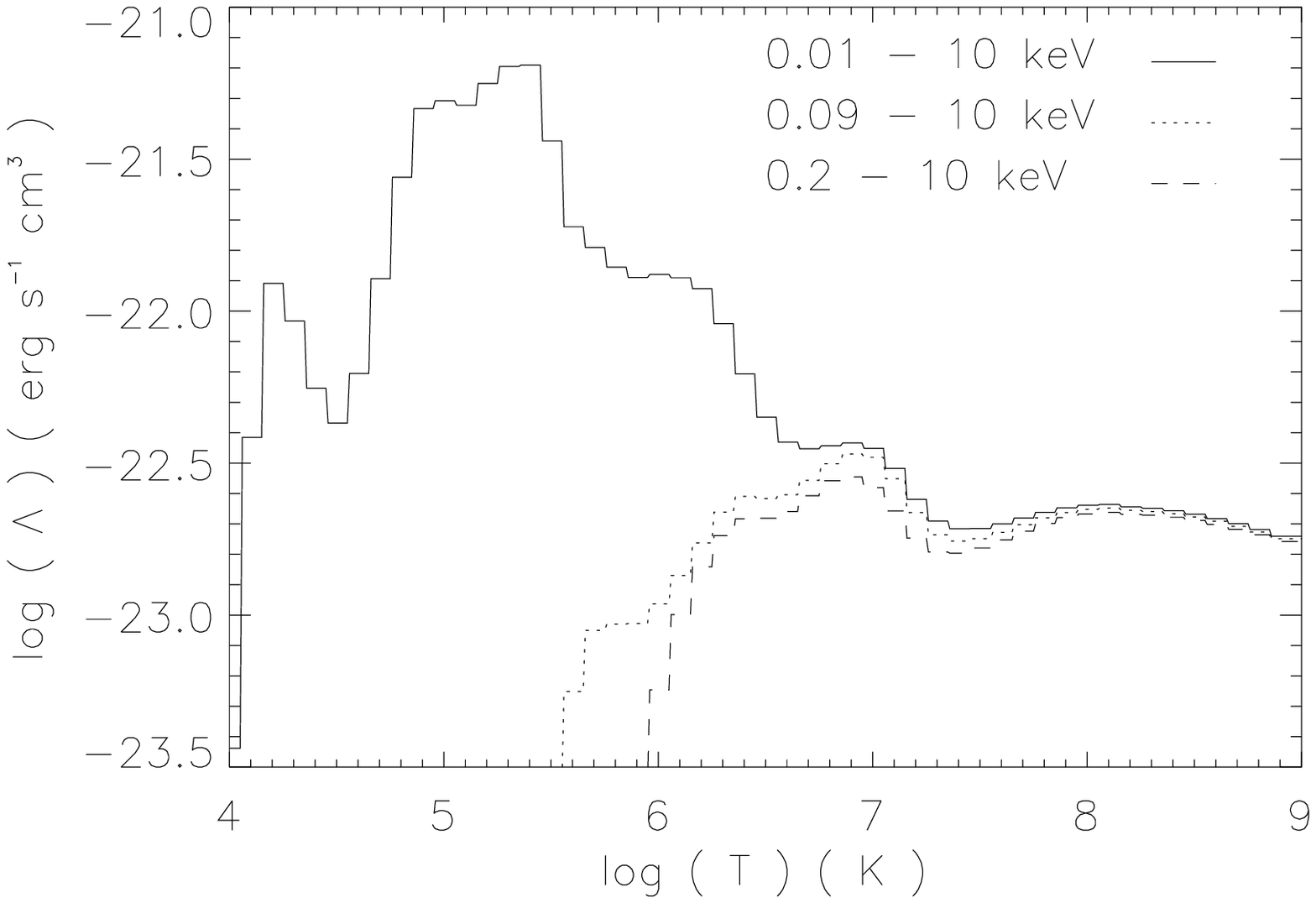}
\plotone{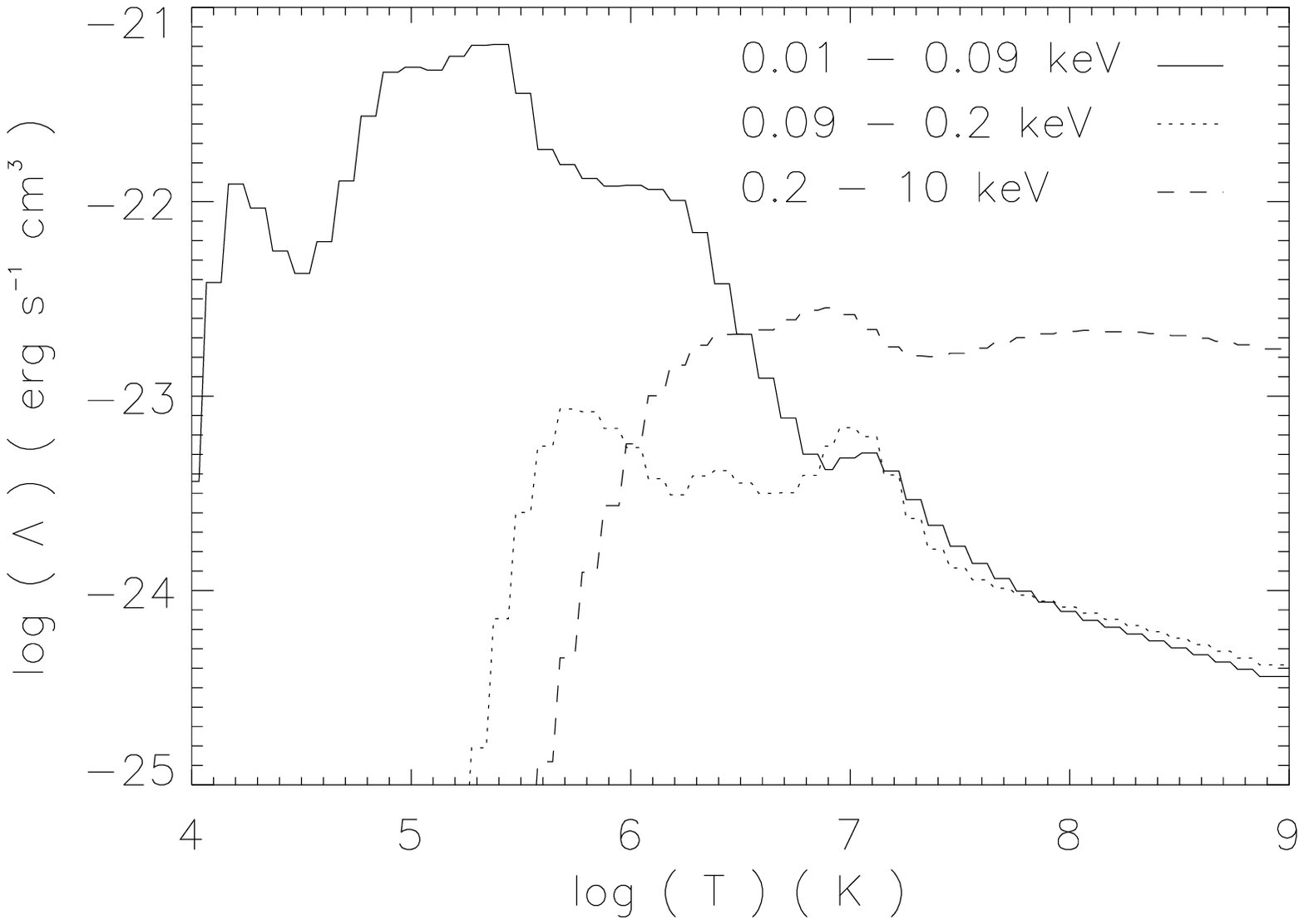}
\caption{Top: X-ray emissivity between 0.01 -- 10 keV (solid), 0.09 -- 10 
keV (dash-dotted) and 0.2 -- 10 keV (dashed) derived from the {\em ATOMDB} as 
a function of the plasma temperature; bottom: X-ray emissivity in the three 
different energy bins 0.01 -- 0.09 keV (solid), 0.09 -- 0.2 keV (dash-dotted) 
and 0.2 -- 10 keV (dashed)}
\label{Xray_emiss}
\end{figure}

X-ray emission in an interacting winds model, in principle, will arise from 
gas with a large range of temperatures above $10^4$ K. We calculate the X-ray 
emission between 0.01 -- 10 keV. The lower energy boundary is determined by 
the limitations of {\em ATOMDB}. We have divided this range into three bins. 
The high energy bin (0.2 -- 10 keV or 1.24 -- 62 \AA) represents the energy 
range covered by the {\em ACIS} instrument and (almost) that of {\em HETG} 
(0.4 -- 10 keV) on CHANDRA. We define a medium energy bin (0.09 -- 0.2 keV or 
62 -- 137 \AA), since the energy ranges of the {\em EPIC}\ instrument on XMM 
(0.1 -- 12 keV) and of {\em LETG} on CHANDRA (0.09 -- 3 keV) extend to lower 
energies, and a low energy bin between 0.01 -- 0.09 keV (137 -- 1180 \AA) 
which is sensitive to gas with temperatures $\lesssim 10^6$ K. In Fig. 
\ref{Xray_emiss}, we show how the X-ray emissivity varies as a function of 
temperature in the three energy bins we have defined above. It is interesting
to note that gas with a temperature lower than $2 \times 10^{6}$ K contributes 
only marginally in the high energy bin which is probed by the {\em ACIS} 
instrument on CHANDRA (Fig. \ref{Xray_emiss}, bottom).

\section{One-dimensional simulations} \label{sec_res_1D}

\subsection{Structure} 
\label{sec_res_1D_str}

\begin{figure}
\plotone{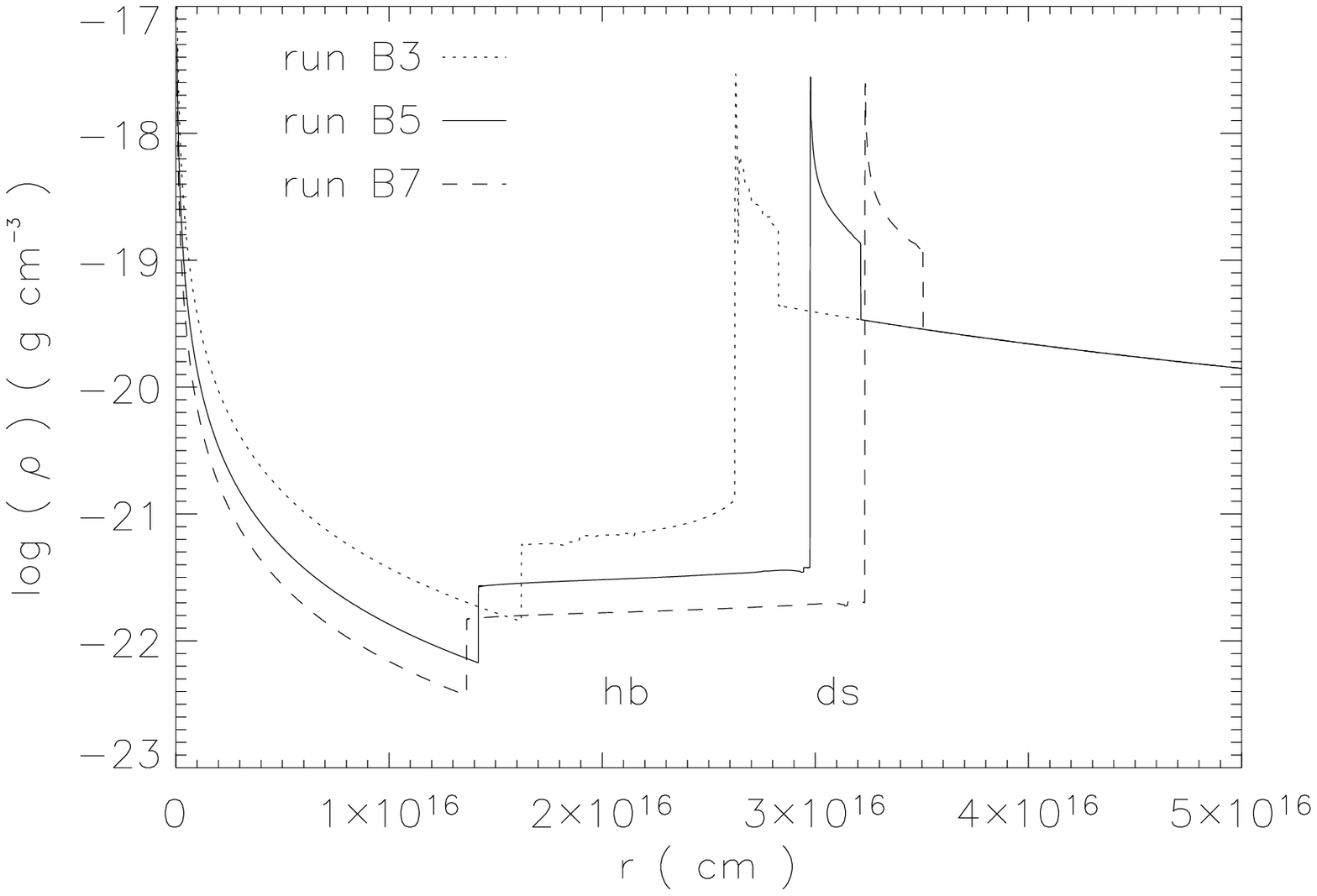}
\plotone{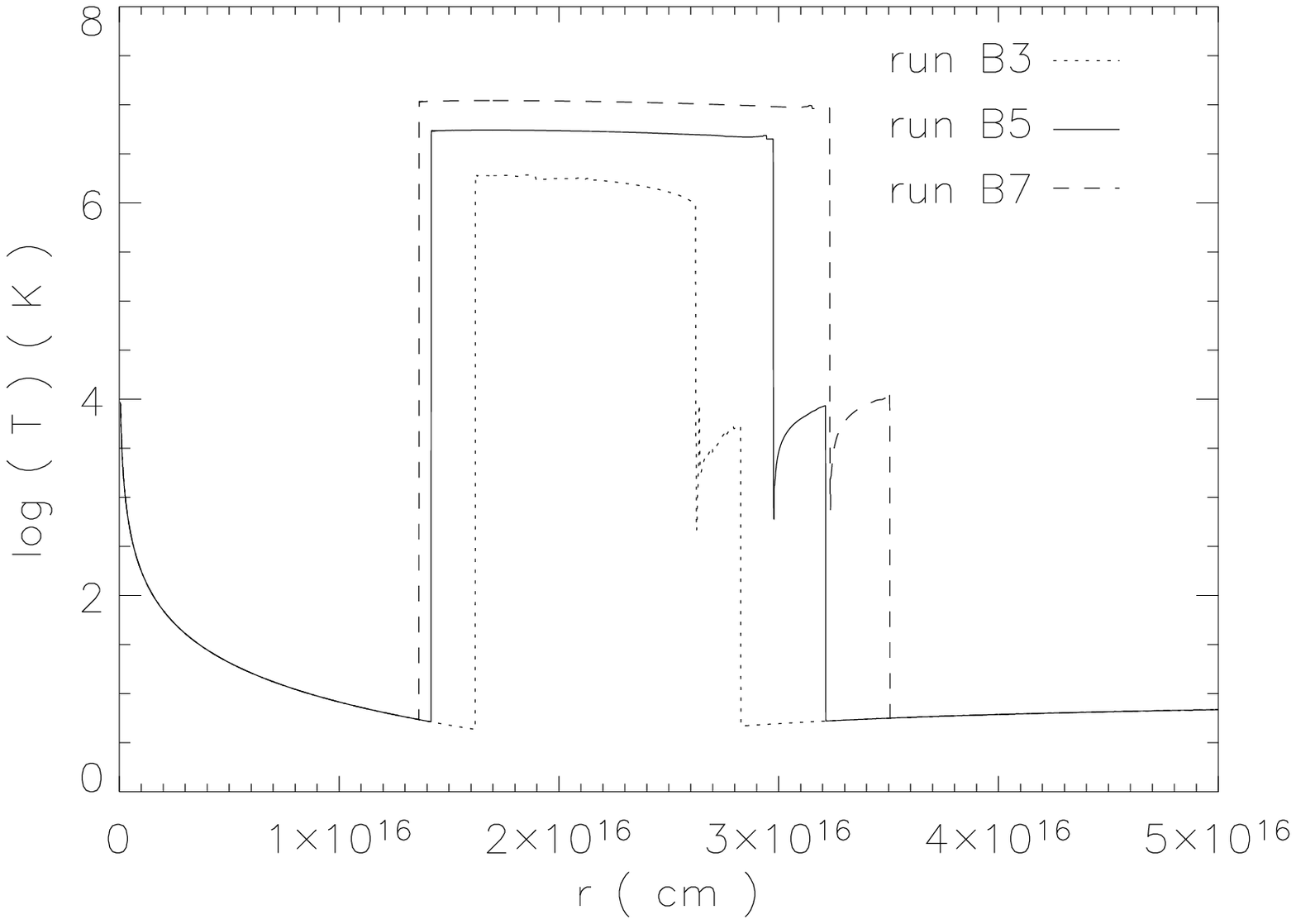}
\caption{Plots of the density and temperature at an age of 390 years in the 
one-dimensional runs B3 (dotted), B5 (solid) and B7 (dashed), 
i.e., $\dot M_{\rm s} = 7 \times10^{-6}$ M$_\odot$ yr$^{-1}$ and $v_{\rm f}$ 
is 300, 500 and 700 km s$^{-1}$, respectively; the position of the dense shell 
(ds) and hot bubble (hb) are indicated for model B5 in the density plot}
\label{1D_vars}
\end{figure}

The profiles of density and temperature as a function of radius (Fig. 
\ref{1D_vars}; runs B3, B5 and B7 at an age of 390 years) show four different 
regions:
(i) the unshocked fast wind, (ii) the shocked fast wind (hot bubble), (iii) 
the shocked slow wind forming a dense shell and (iv) the unshocked slow wind. 
The reverse shock separates region (i) from (ii). The large jump in density 
and temperature between regions (ii) and (iii) represents the contact 
discontinuity. The outer shock separates region (iii) from (iv). Due to the 
interaction of the fast and the slow wind, both regions of shocked matter are 
heated to high temperatures, but the dense shell is highly radiative and cools 
quickly.
The evolution of the shell normally has the following stages: first there is
free expansion, if cooling sets in there is then the radiative,
momentum-driven bubble and finally the adiabatic, energy-driven bubble.

Depending on the velocity of the fast wind, namely if it is larger or smaller 
than a critical velocity 
\begin{eqnarray} \label{eq_vcr}
v_{\rm cr} &=& \left( \frac{\mathcal{L}}{6\,\pi\,\rho_{01}} \right)^{1/3} = 
\left( \frac{1}{3}\,\frac{\dot M_{\rm f}\,v_{\rm f}^2
\,v_{\rm s}}{\dot M_{\rm s}} \right)^{1/3} ,
\end{eqnarray}
the end stage is then the momentum-driven or energy-driven bubble 
\citep{KMK92a,KMK92b}, where $\mathcal{L} = 1/2\,\dot M_{\rm f}\,v_{\rm f}^2$ 
is the kinetic luminosity of the fast wind. This critical velocity is 
independent of the cooling function in the case of a constant mass outflow rate
of the fast wind and a $r^{-2}$ density profile \citep{KMK92b}. As the fast 
wind is always faster than $v_{\rm cr}$ in our models, the bubble is expected 
to go from free expansion phase directly to the energy-driven phase. This 
transition occurs in our models within a short period of time (about few 
years), after which the velocity of the contact discontinuity reaches a stable 
value characteristic of an energy-driven shell. In the momentum-driven case, 
the position of the shell would be \citep{KMK92a,KMK92b} 
\begin{eqnarray}
R_{\rm cd}^{\rm M} &=& \frac{\mathcal{I}}{1 + \mathcal{I}}\,v_{\rm f}\,t 
\approx \left( \frac{\dot M_{\rm f}\,v_{\rm f}\,v_{\rm s}}{\dot M_{\rm s}} 
\right)^{1/2} \,t ,
\end{eqnarray}
with  
\begin{equation}
\mathcal{I} = \left( \frac{\mathcal{L}}{2\,\pi\,\rho_{01}\,v_{\rm f}^3} 
\right)^{1/2} = \left( \frac{\dot M_{\rm f}\,v_{\rm s}}{\dot M_{\rm s}\,
v_{\rm f}} \right)^{1/2} .
\end{equation}
The resulting velocity of about 10 km s$^{-1}$ (Table 1) are significantly 
lower than those found in our simulations. In the energy-driven, adiabatic 
case, the analytical solutions of the position of the shell, $R_{\rm cd}$, is 
\begin{eqnarray}  \label{eq_vcde}
R_{\rm cd}^{\rm E} &=& \left( \frac{\Gamma_{\rm rad}\,\xi\,\mathcal{L}}{3\,
\rho_{01}} \right)^{1/3} \,t \nonumber \\
&=& \left( \frac{2\,\pi\,\Gamma_{\rm rad}\,\xi\,\dot M_{\rm f}\,v_{\rm f}^2\,
v_{\rm s}}{3\,\dot M_{\rm s}} \right)^{1/3} \,t ,
\end{eqnarray}
with $4\,\pi\,\Gamma_{\rm rad}\,\xi = 4.165$ \citep{KMK92a,KMK92b}. Here the 
ambient shock is assumed to be adiabatic. Assuming a radiative ambient shock, 
$4\,\pi\,\Gamma_{\rm rad}\,\xi = 2.0$ and the velocity of the shell reduces 
to $v_{\rm cr}$. $\Gamma_{\rm rad}$ is the fraction of the injected energy in 
the bubble which is radiated away, $\xi$ is a numerical constant of order 
unity \citep{KMK92b}. The velocity of the contact discontinuity 
$v_{\rm cd, sim}$ in our simulations is within the range of velocities 
between the limiting values given by the radiative and adiabatic 
approximations -- $v_{\rm cr}$ and $R_{\rm cd}^{\rm E}/t$ -- for the dense 
shell of \citet{KMK92a,KMK92b} (see Table 1). 

The density and temperature of the hot bubble are in good agreement with 
results from the self-similar formulation of \citet{ChI83}, although radiative 
cooling emphasizes the density and temperature jumps close to the contact 
discontinuity. For example in our model B5, the density is about 
$3 \times 10^{-22}$ g cm$^{-3}$, i.e. a number density of about 200 cm$^{-3}$, 
and the temperature is $5 \times 10^6$ K. Our density is slightly lower than 
the value derived from the self-similar formulation of 350 cm$^{-3}$ -- this 
leads to a reduced cooling efficiency, and thus our temperature is slightly 
higher than the value from the self-similar model of $3.4 \times 10^6$ K. 

The temperature of the hot bubble is given in Table 2. We calculated an X-ray 
weighted average temperature as defined in ASB06. It is only a 
function of the velocity of the fast wind, ranging from $1.5 \times 10^6$ K 
for $v_{\rm f} = 300$ km s$^{-1}$ to $2.2\times10^7$ K for $v_{\rm f} = 1000$ 
km s$^{-1}$. The mass loss rate of the slow wind is fairly unimportant, i.e., 
the temperature is almost identical, e.g., in the models A3, B3 and C3. The 
temperature is roughly given by
\begin{equation} \label{eq_rh}
T_{\rm hb} = 1.4\times10^6\,{\textrm K}\,\left( \frac{v_{\rm f}}{300\,{\rm km}
\,{\rm s}^{-1}} \right)^2
\end{equation}
in which the dependence on only $v_{\rm f}$ is as expected from the 
Rankine-Hugoniot conditions \citep[e.g.][]{LaL59}. As the bubble evolves, 
the temperature of the hot bubble remains fairly constant with time.

\subsection{X-ray properties} \label{sec_res_1D_xray}

We now present the properties of the X-ray emission from our models and their 
dependence on the physical parameters of the fast wind. 

\subsubsection{The X-ray luminosities} \label{sec_res_1D_xray_lum}

\begin{figure}
\plotone{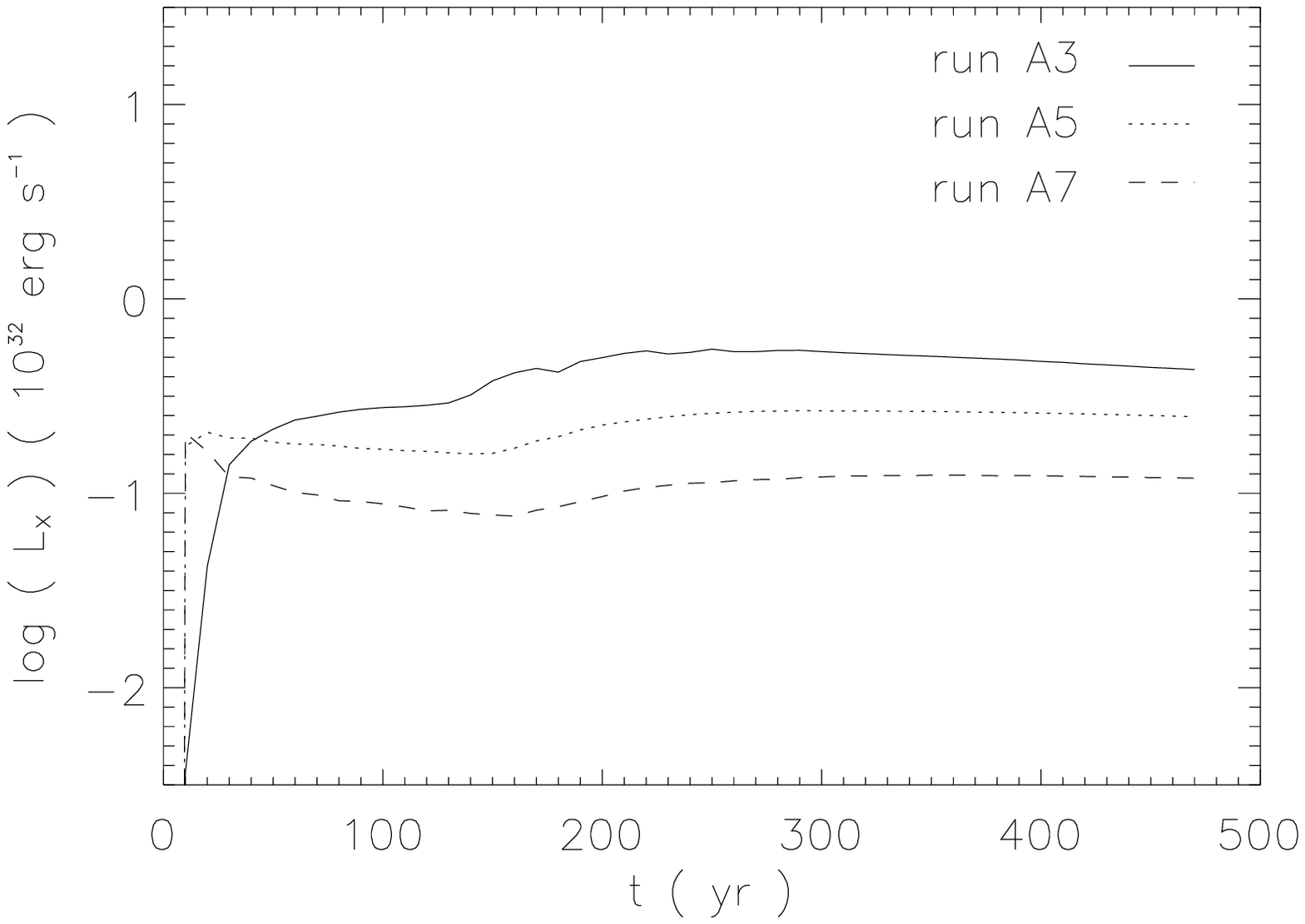}
\plotone{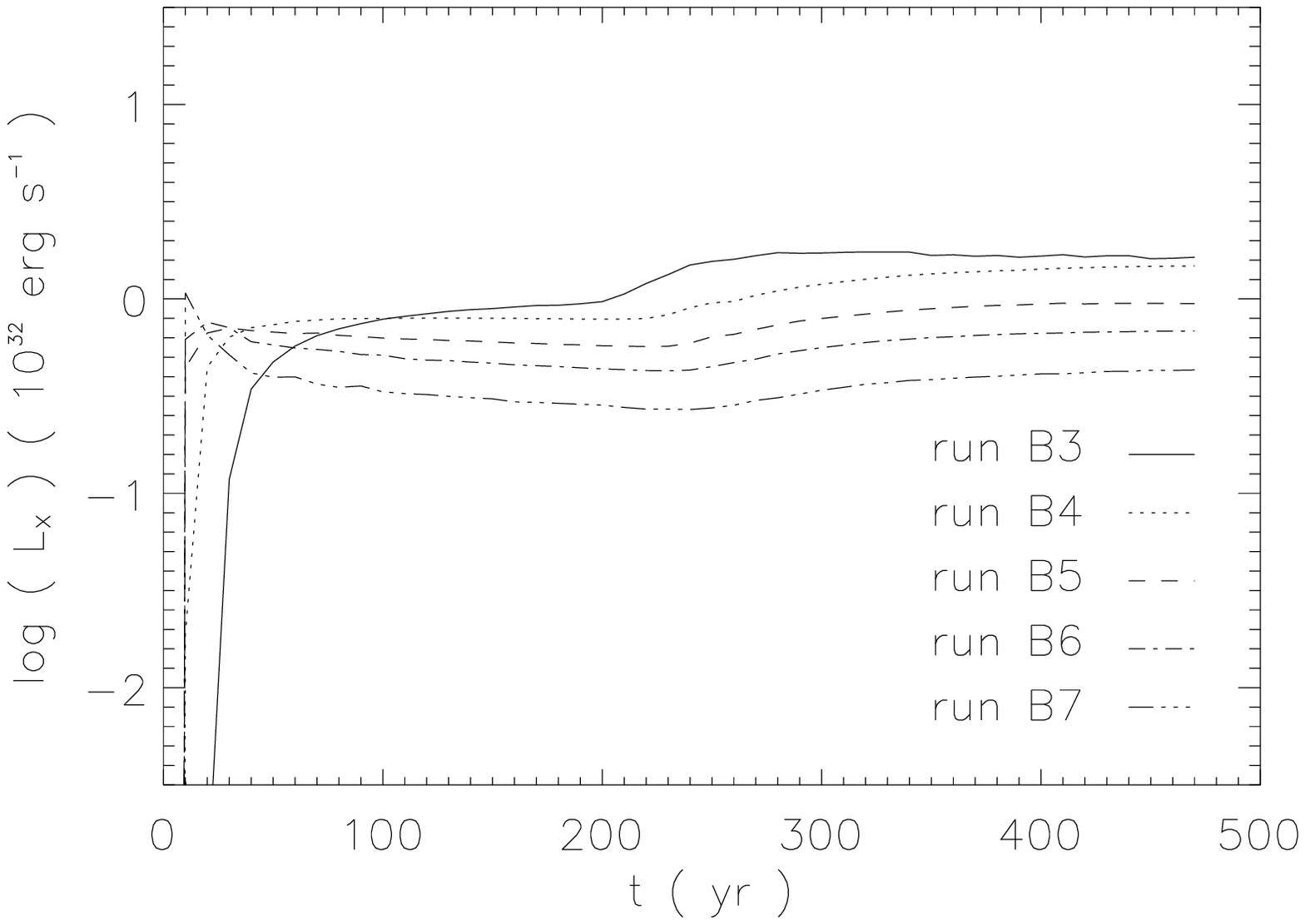}
\plotone{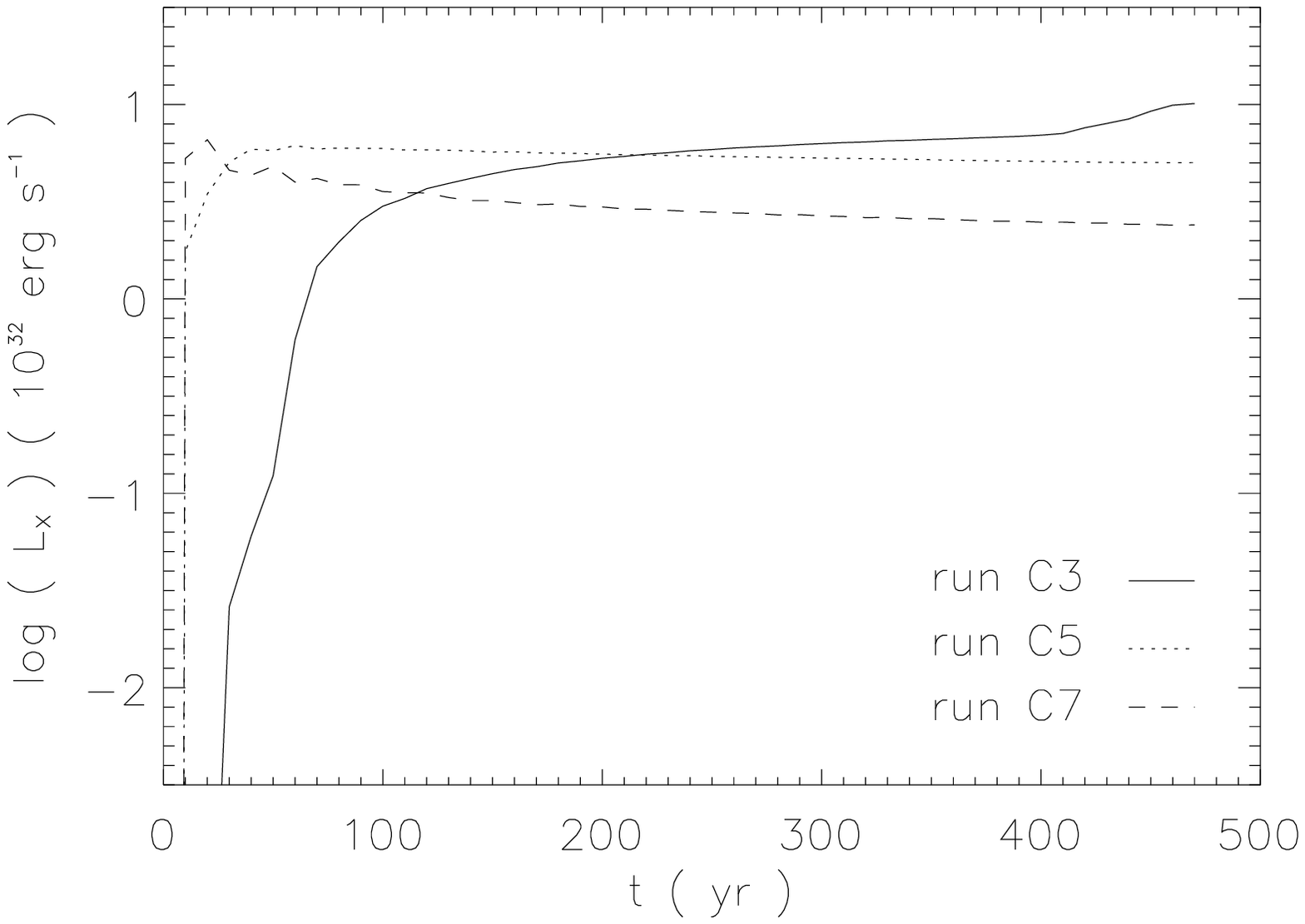}
\caption{X-ray luminosity $L_{\rm x,ACIS}$ in the high energy bin between 
0.2 -- 10 keV as a function of the evolution time in the 
one-dimensional runs A3-A7 (top), B3-B7 (middle) and C3-C7 (bottom)} 
\label{Lvst_allruns}
\end{figure}

Several clear trends in the X-ray luminosity are visible between the different 
sets of simulations. We show the X-ray luminosity in the high energy bin 
between 0.2 -- 10 keV, $L_{\rm x,ACIS}$, as a function of the evolution time 
for all models in Fig. \ref{Lvst_allruns}. Comparing models from different 
sets with fixed $\dot M_{\rm s}$ which have the same fast wind velocity, we 
find that the higher the mass outflow rate of the fast wind is, the higher 
$L_{\rm x,ACIS}$ is. For example, at an age of 470 years, in runs A3, B3 and 
C3 $L_{\rm x,ACIS}$ is $0.1 \times 10^{32}$ erg s$^{-1}$, $0.4\times 10^{32}$ 
erg s$^{-1}$ and $2.4 \times 10^{32}$ erg s$^{-1}$, respectively. In these 
runs, the fast wind velocity is 300 km s$^{-1}$ and $\dot M_{\rm f}$ is 
$10^{-7}$, $2.33\times 10^{-7}$ and $10^{-6}$ M$_\odot$ yr$^{-1}$, 
respectively. 

Comparing the curves within each panel (each panel represents a fixed 
$\dot M_{\rm s}$ and thus a constant $\dot M_{\rm f}\,v_{\rm f}$), 
$L_{\rm x,ACIS}$ decreases with increasing fast wind velocity. This is because 
the correspondingly lower mass outflow rate of the fast wind\footnote{A result 
of the assumption $\dot M_{\rm f}\,v_{\rm f} = \dot M_{\rm s}\,v_{\rm s}$}
leads to lower densities in the shocked fast wind and therefore to reduced 
X-ray emission. 

\begin{deluxetable*}{cccccc}
\tablehead{\colhead{run} & \colhead{$R_{\rm cd}$ ( 10$^{16}$ cm )} & 
\colhead{$R_{\rm cd} / v_{\rm ASB}$ ( yrs )} & \colhead{$L_{\rm ASB}$ ( erg 
s$^{-1}$ )} & \colhead{$L_{\rm x,ACIS}$ ( erg s$^{-1}$ )} & 
\colhead{$T_{\rm hb}$ ( K )}}
\startdata
A3 & 3.4 & 377 & $1.3\times10^{31}$ & $4.33\times10^{31}$ & $1.5\times10^{6}$\\
A5 & 3.9 & 379 & $2.\times10^{31}$ & $2.48\times10^{31}$ & $4.5\times10^{6}$\\
A7 & 4.3 & 383 & $1.3\times10^{31}$ & $1.20\times10^{31}$ & $8.8\times10^{6}$\\
\tableline
B3 & 3.2 & 355 & $6.9\times10^{31}$ & $1.64\times10^{32}$ & $1.5\times10^{6}$\\
B4 & 3.5 & 360 & -- & $1.48\times10^{32}$ & $3.0\times10^{6}$\\
B5 & 3.7 & 361 & $1.1\times10^{32}$ & $9.48\times10^{31}$ & $4.9\times10^{6}$\\
B6 & 3.8 & 352 & -- & $6.84\times10^{31}$ & $7.1\times10^{6}$\\
B7 & 4.0 & 356 & $7.6\times10^{31}$ & $4.31\times10^{31}$ & $9.6\times10^{6}$\\
\tableline
C3 & 3.0 & 332 & $<2.\times10^{31}$ & $1.01\times10^{33}$ & $1.7\times10^{6}$\\
C5 & 3.4 & 330 & $1.4\times10^{33}$ & $5.03\times10^{32}$ & $5.3\times10^{6}$\\
C7 & 3.7 & 329 & $1.3\times10^{33}$ & $2.40\times10^{32}$ & $1.1\times10^{7}$\\
\tableline
A10 & 4.7 & -- & -- & $2.84\times10^{30}$ & $1.8\times10^{7}$\\
B10 & 4.4 & -- & -- & $1.00\times10^{31}$ & $2.0\times10^{7}$\\
D7 & 4.0 & -- & -- & $9.05\times10^{32}$ & $1.1\times10^{7}$\\
E10 & 3.0 & -- & -- & $2.34\times10^{33}$ & $2.2\times10^{7}$\\
\enddata
\tablecomments{The X-ray luminosity $L_{\rm x,ACIS}$ in the 
high-energy bin 0.2 -- 10 keV at an age of 470 years in our simulations and 
results 
$L_{\rm ASB}$ from ASB06 (from epochs where the bubbles have equal size in our 
models and those of ASB06, since the velocity of the contact discontinuity is 
slightly different in our models compared to ASB06); X-ray weighted average
temperature of the hot bubble $T_{\rm hb}$ as defined in ASB06 at an age of 
470 years}
\end{deluxetable*}
We can also investigate the dependence of $L_{\rm x,ACIS}$ on $v_{\rm f}$, when
$\dot M_{\rm f}$ is constant. Using pairs of our runs, as e.g. runs A3 and B7 
(in which $\dot M_{\rm f} = 10^{-7}$ M$_\odot$ yr$^{-1}$) or C3 and D7 
($\dot M_{\rm f} = 10^{-6}$ M$_\odot$ yr$^{-1}$), we can see that the 
luminosities are almost the same in each pair (Table 2), even though, due to 
the different velocities, the temperatures in the hot bubble are very 
different. This is a somewhat surprising result, but which can be understood 
in terms of the important physical quantities determining the X-ray luminosity.
These are the volume, density and temperature of the hot bubble. The volumes 
of the hot bubble in run C3 and D7 are about $6.84 \times 10^{49}$ cm$^3$ and 
$2.2 \times 10^{50}$ cm$^3$, respectively (derived from 
$R_{\rm cd}=3.0 \times 10^{16}$ cm and $R_{\rm rs}=2.2 \times 10^{16}$ cm for  
C3 and $R_{\rm cd}=3.9 \times 10^{16}$ cm and $R_{\rm rs}=2.2 \times 10^{16}$ 
cm for D7). The mean (radially averaged) density in the hot bubble for C3 (D7) 
is $1.66 \times 10^{-21}$ g cm$^{-3}$ ($6.43 \times 10^{-22}$ g cm$^{-3}$), 
the emissivity at the mean temperature of the hot bubble -- $1.73 \times 10^6$ 
K ($1.06 \times 10^7$ K) -- is $1.44 \times 10^{-23}$ erg s$^{-1}$ cm$^3$ 
($2.63 \times 10^{-23}$ erg s$^{-1}$ cm$^3$), thus the luminosity is 
$9.7 \times 10^{32}$ erg s$^{-1}$ ($8.6 \times 10^{32}$ erg s$^{-1}$). These 
estimates are in good agreement with the more accurate values of the luminosity
in Table 2. In the other pair A3 and B7, the differences between the estimates 
using radially averaged quantities and the values in Table 2 are somewhat 
larger -- the luminosities in runs A3 and B7 are $4.9 \times 10^{32}$ erg 
s$^{-1}$ and $3.9 \times 10^{32}$ erg s$^{-1}$, respectively, compared to 
$4.3 \times 10^{32}$ erg s$^{-1}$ in both cases in Table 2. This is because 
the density and temperature gradients across the hot bubble are larger than in 
the first pair and therefore also the errors, we introduce by the averaging 
process.

\begin{figure}
\plotone{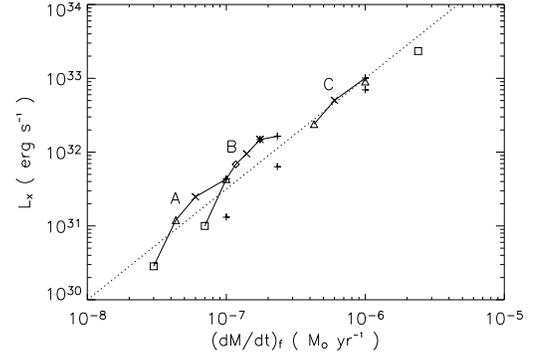}
\caption{X-ray luminosity $L_{\rm x,ACIS}$ in the high energy bin between 
0.2 -- 10 keV as a function of the mass loss rate of the fast wind 
$\dot M_{\rm f}$ for all computed models at an age of 470 years (plus sign:
$v_{\rm f} = 300$ km 
s$^{-1}$, asterisk: 400 km s$^{-1}$, cross: 500 km s$^{-1}$, diamond: 600 km 
s$^{-1}$, triangle: 700 km s$^{-1}$, square: 1000 km s$^{-1}$); the dotted line
represents the curve $L_{\rm x,ACIS} \sim \dot M_{\rm f}^{3/2}$. In all 
models except for run E10 (the model with the highest $\dot M_{\rm f}$) we 
assumed $\dot M_{\rm f}\,v_{\rm f} = \dot M_{\rm s}\,v_{\rm s}$. 
The solid lines join models within each set of constant 
$\dot M_{\rm s}$. The relation appears to hold roughly for ages of PNs up to 
$\sim$2000 years (unjoined plus signs: runs A3, B3, and C3 at an age of 1900 
years).}
\label{Lvsmdotf}
\end{figure}

In summary, the weak dependence of the X-ray luminosity, $L_{\rm x,ACIS}$, on 
$v_{\rm f}$ has two reasons: i) the emissivity in the high energy bin is 
fairly flat over the range of temperatures in our 
models (Fig. \ref{Xray_emiss}) and 
ii) the volume of the hot bubble depends weakly on $v_{\rm f}$  
($R_{\rm cd} \sim v_{\rm f}^{1/3}$) due to our assumption of 
$\dot M_{\rm f}\,v_{\rm f} = \dot M_{\rm s}\,v_{\rm s}$. In Fig. 
\ref{Lvsmdotf}, we illustrate that $L_{\rm x,ACIS}$ depends much more strongly 
on $\dot M_{\rm f}$ than on $v_{\rm f}$ by plotting the values for all 
computed models (therefore the full range of $v_{\rm f}$). The data can be 
reasonably well fitted by the curve $L_{\rm x,ACIS} \sim \dot M_{\rm f}^{3/2}$ 
-- the scatter of the individual data points above and below this curve lies 
within a factor of 2. Interestingly, even model E10 where 
$\dot M_{\rm f}\,v_{\rm f} = 0.5 \dot M_{\rm s}\,v_{\rm s}$ fits the 
$L_{\rm x,ACIS} \sim \dot M_{\rm f}^{3/2}$ relationship.

We now show that this variation of $L_{\rm x,ACIS}$ with $M_{\rm f}$ can be 
partially understood by considering the dependencies of the volume ($V$),
temperature ($T_{\rm hb}$) and density of the hot bubble ($\rho_{\rm hb}$) on 
$\dot M_{\rm f}$ and $v_{\rm f}$ across and within the different sets of 
models. We utilize analytical estimates of these dependencies based on the 
self-similar solutions without radiative cooling (see \S \ref{sec_res_1D_str}),
therefore we expect some differences between the dependencies derived below 
and those actually found from our modeling (which takes radiative 
cooling effects into account).
The volume of the hot bubble depends on the radii of the reverse shock 
($R_{\rm rs} = [ 3\, \dot M_{\rm f}\,v_{\rm f}\,v_{\rm s} / 
( 4\,\dot M_{\rm s} ) ]^{1/2}\,t = \sqrt{3 / 4}\,v_{\rm s}\,t$) and 
the contact discontinuity (given by eqn. \ref{eq_vcr}). In the range of 
$T_{\rm hb}$ in our models ($1.5 \times 10^6$ -- $2 \times 10^7$ K),
the emissivity changes only within a factor of 1.5 (Fig. \ref{Xray_emiss}) 
and therefore we assume it to be constant. The density of the hot bubble can be
approximated as constant with radius (Fig. \ref{1D_vars}) and equal to the 
density of the fast wind given by eqn. (\ref{eq_rho_CFW}) at the reverse shock 
scaled by a constant compression factor of 4, valid for an adiabatic shock. 

In general, $L_{\rm x,ACIS}$ is a function of the 4 parameters in our models 
-- $v_{\rm f}$, $\dot M_{\rm f}$, $v_{\rm s}$ and $\dot M_{\rm s}$. Since 
$v_{\rm s}$ is kept constant in all our models and we assume 
$\dot M_{\rm f}\,v_{\rm f} = \dot M_{\rm s}\,v_{\rm s}$, the number of 
independent parameters is only two -- we select $v_{\rm f}$ and 
$\dot M_{\rm s}$. By holding each of these constant in turn one 
can get insights into the behavior of $L_{\rm x,ACIS}$ as a function of 
$\dot M_{\rm f}$. 

First we keep $v_{\rm f}$ constant, i.e. we compare models from different sets 
with the same $v_{\rm f}$. Then the radius of the contact discontinuity is
constant, thus also the volume. Hence $\rho_{\rm hb} \sim \dot M_{\rm f}$ 
and $\rho_{\rm hb}^2\,V \sim \dot M_{\rm f}^2$. Therefore we expect 
$L_{\rm x,ACIS}$ to vary as $\dot M_{\rm f}^2$. In our models, the runs 
with the same $v_{\rm f}$ follow $L_{\rm x,ACIS} \sim \dot M_{\rm f}^{1.35}$, 
which is primarily caused by a departure from the analytical estimate of the 
model (radially-averaged) density dependency, $< \rho > \sim 
\dot M_{\rm f}^{0.75}$.  

Next we keep $\dot M_{\rm s}$ constant, i.e. we compare models within each set,
thus $v_{\rm f}\,\dot M_{\rm f}$ is constant. Hence $R_{\rm cd} \sim 
v_{\rm f}^{1/3} \sim \dot M_{\rm f}^{-1/3}$ and therefore 
$V \sim \dot M_{\rm f}^{-1}$. The density $\rho_{\rm hb} \sim 
\dot M_{\rm f}^2$ and therefore $\rho_{\rm hb}^2\,V \sim \dot M_{\rm f}^3$, 
implying $L_{\rm x,ACIS} \sim \dot M_{\rm f}^3$. This expectation is supported 
by the steep increase in $L_{\rm x,ACIS}$ with $\dot M_{\rm f}$ which we see 
in sets A and B for models with the lower values of $\dot M_{\rm f}$, i.e. 
with the larger values of $v_{\rm f}$ (500 to 1000 km s$^{-1}$). For our 
models with lower values of $v_{\rm f}$, however, the increase is not as steep,
due to two effects: (i) deviations from the analytical dependencies as well as 
(ii) the replacement of the radial integral for $L_{\rm x,ACIS}$ with the 
product of radial-averages.
For example, as a result of effect (i) in our models B3 and B4 we find 
$R_{\rm rs} \sim \dot M_{\rm f}^{0.35}$, $R_{\rm cd} 
\sim \dot M_{\rm f}^{-0.28}$, $V \sim \dot M_{\rm f}^{-1.21}$, 
$\rho_{\rm hb} \sim \dot M_{\rm f}^{1.84}$. Furthermore the emissivity in run 
B3 in 30 \% lower compared to that in B4, equivalent to $\Lambda \sim 
\dot M_{\rm f}^{-0.9}$. The net result is that $L_{\rm x,ACIS} \sim 
\rho_{\rm hb}^2\,V\,\Lambda \sim \dot M_{\rm f}^{1.6}$. As a result of effect
(ii), the radial integral becames increasingly overestimated by the product of 
radially-averaged quantities for models with the lowest values of $v_{\rm f}$,
where the radial gradients of these quantities are the largest 
(e.g. compare run B3 and B5 in Fig. \ref{1D_vars}). Hence the expected 
dependence of $L_{\rm x,ACIS}$ on $\dot M_{\rm f}$ will be even shallower as
observed.

The X-ray luminosities in our different models do not change strongly with 
time as the bubble evolves, after an initial phase of increasing luminosity
(e.g. a period of $\sim 100$ years in run C3). The length of the initial phase 
of rising luminosity is primarily given by the cooling time in the hot bubble. 
Once a gas parcels has cooled below $10^6$ K, it does not contribute to the 
luminosity $L_{\rm x,ACIS}$ of the hot bubble anymore. Within each set, the 
lower the value of $v_{\rm f}$, the lower the temperature and the higher the 
density in the hot bubble, thus the shorter the cooling time. Hence in these 
case, the length of the initial phase is the highest. We discuss the weak 
dependence of $L_{\rm x,ACIS}$ on time and the initial phase of rising 
luminosity and compare our results with those of ASB06 in \S \ref{sec_diff}.
The relation $L_{\rm x,ACIS} \sim \dot M_{\rm f}^{3/2}$ shown in 
Fig. \ref{Lvsmdotf} appears to hold for ages of PNs up to $\sim$2000 years, 
presumably because of the similar weak decay of $L_{\rm x,ACIS}$ with time in 
all models.

\begin{figure}
\plotone{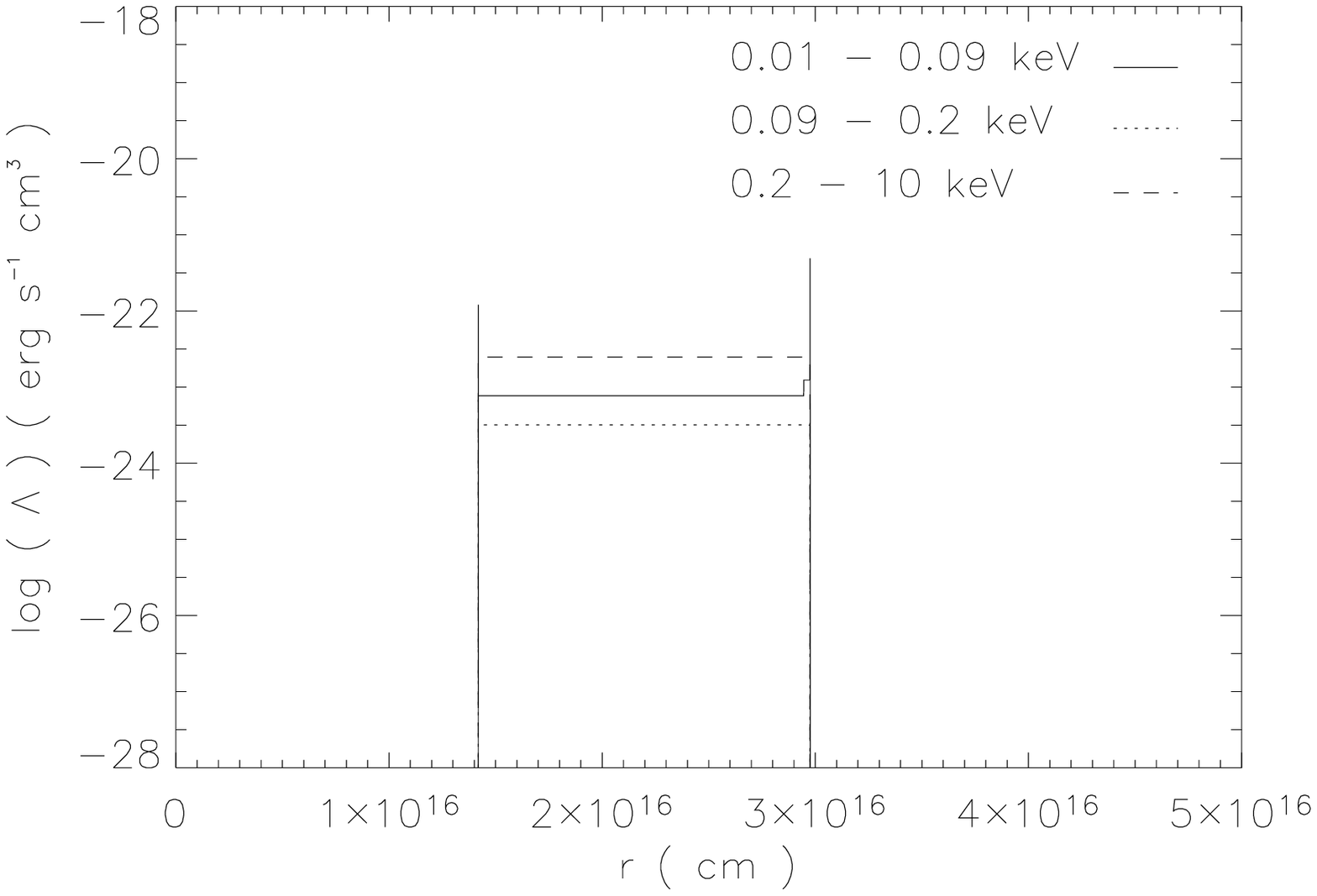}
\plotone{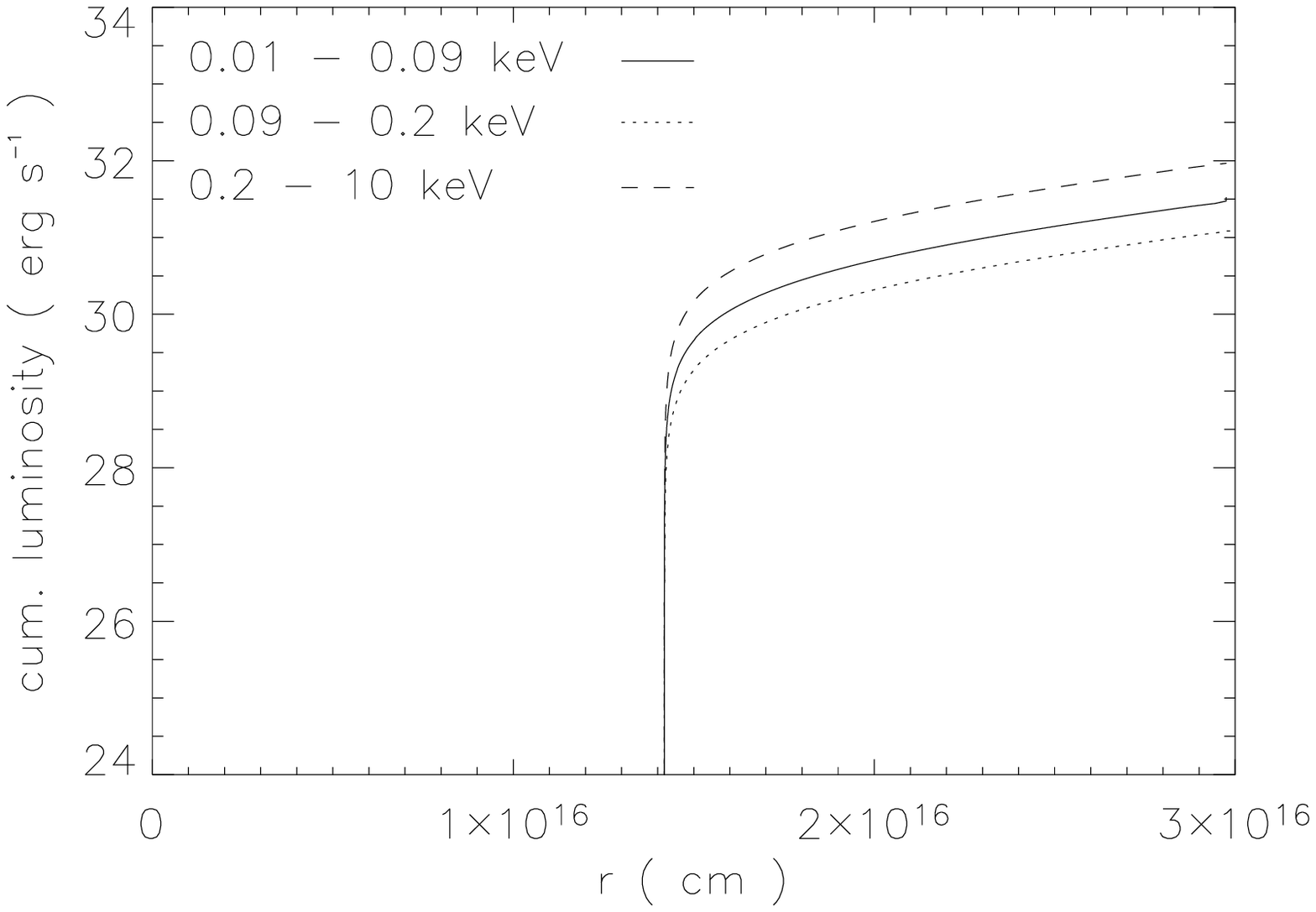}
\caption{Top: Plot of the X-ray emissivity in the three energy bins 0.01 --
0.09 keV (solid), 0.09 -- 0.2 keV (dotted) and 0.2 -- 10 keV (dashed) as a 
function of radius at an age of 390 years of the one-dimensional run B5; 
bottom: Cumulative plot of the luminosity inside a given radius as a function 
of this radius}
\label{1D_spec}
\end{figure}

Up to now, we only took into account luminosities in the high energy bin 
0.2 -- 10 keV, which can be observed with the {\em ACIS} instrument. An 
inspection of the X-ray luminosity emitted in the lower energy bins in our 
models may be of potential use in identifying new observational probes of 
cooler gas in the hot bubble near its interface with the dense shell. Plotting 
the X-ray emissivity and luminosity in our three energy bins (e.g. for run 
B5), we find that most of the emission from the main body of the hot bubble, 
i.e. excluding its edges, is radiated in the high energy bin. However, the 
total energy radiated by the main body of the hot bubble in these three energy 
bins $L_{\rm hb, m}$ ($1.4 \times 10^{32}$ erg s$^{-1}$) is significantly 
smaller than the rate at which mechanical energy is pumped into the system by 
the fast wind, $L_{\rm fw}=1.1 \times 10^{34}$ erg s$^{-1}$. This mechanical 
energy is partly converted into the mechanical energy of the dense shell, 
which increases at a rate given by 
$L_{\rm ds} = 1/2\,\dot M_{\rm s}\,v_{\rm cd}^3 / v_{\rm s} \approx
4.6 \times 10^{33}$ erg s$^{-1}$, and the remaining energy goes into the 
hot bubble. The energy input into the hot bubble $L_{\rm hb}$ 
is about $L_{\rm fw} - L_{\rm ds} = 6 \times 10^{33}$ erg s$^{-1}$, from of 
which $L_{\rm ad} = 1/3\,\dot M_{\rm s}\,v_{\rm cd}^3 / v_{\rm s} \approx 3 
\times 10^{33}$ erg s$^{-1}$ is required for the adiabatic expansion of the hot
bubble. Therefore the energy radiated by the hot bubble,
$L_{\rm hb} - L_{\rm ad} = 3 \times 10^{33}$ erg s$^{-1}$, is significantly 
larger than $L_{\rm hb, m}$ and this difference  must be radiated by the 
edges. This is qualitatively demonstrated by a jump in the cumulative 
luminosity function, $C_{\rm x} ( r )$ (i.e., the luminosity emitted by 
material inside a given radius $r$), across the contact discontinuity. There 
are spikes in the emissivity in the low energy bin at both 
edges of the hot bubble (Fig. \ref{1D_spec}, top) due to the presence of gas 
with intermediate temperatures $\lesssim 10^{5.5}$ K. However, only the outer 
edge produces a jump in $C_{\rm x} ( r )$ for the low energy bin, because it 
has a significantly higher emission measure (see Fig. \ref{1D_vars}, top).

Hence the maximum fractional contribution to the X-ray luminosity emitted in 
the full energy range (0.01 -- 10 keV) lies in the low energy bin and comes, 
not from the main body of the hot ($5\times 10^6$ K) bubble, but from its 
somewhat cooler ($10^4$ -- $10^6$ K) edges. The gas in the dense shell 
is too cool ($< 10^4$ K) to emit any X-ray radiation. The structure of the hot 
bubble is similar in the high and medium energy bins, however, it is 
significantly limb-brightened in the low energy bin.

\begin{figure}
\plotone{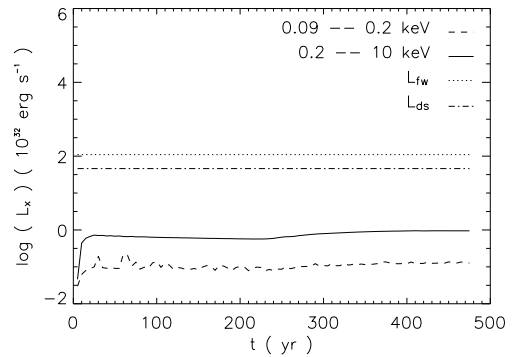}
\caption{X-ray luminosity as a function of the evolution time 
in the one-dimensional run B5 in the energy bins 0.09 -- 0.2 keV (dashed) and 
0.2 -- 10 keV (solid); also plotted are the mechanical energy injected per 
unit time by the fast wind (dotted) and the increase of mechanical energy per 
unit time of the dense shell (dash-dotted)}
\label{runB5_1D_Lvst}
\end{figure}

A plot of the luminosity in the high and medium energy bins as a function of 
the evolution time (Fig. \ref{runB5_1D_Lvst}) shows that these are roughly 
constant with time. Note, that since most of the energy in the 0.01 -- 0.09 
keV range is emitted by only a few grid cells in our models, we do not have 
reliable quantitative estimates of the luminosity in the low energy bin.

\subsubsection{The X-ray spectra}

\begin{figure}
\plotone{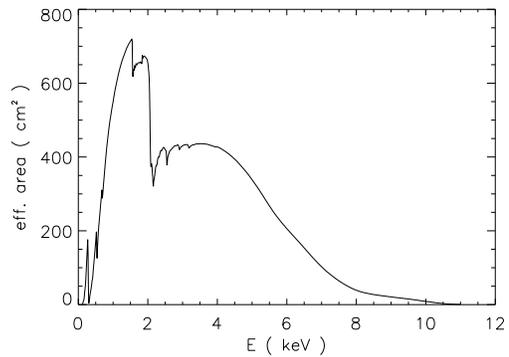}
\caption{Effective area of the {\em ACIS} S instrument on CHANDRA} 
\label{area_ACIS}
\end{figure}

In this section, we investigate the X-ray spectra from our models and how
they are affected by different physical parameters. For this 
purpose, the model spectra have been convolved with the effective area of the 
{\em ACIS} S instrument (Fig. \ref{area_ACIS}). We assume a distance of 1 kpc. 
Our spectra are presented with a resolution of 0.01 keV. The default abundances
in {\em ATOMDB} (see \S \ref{sec_calc_xray}) are used.
\begin{figure*}
\plottwo{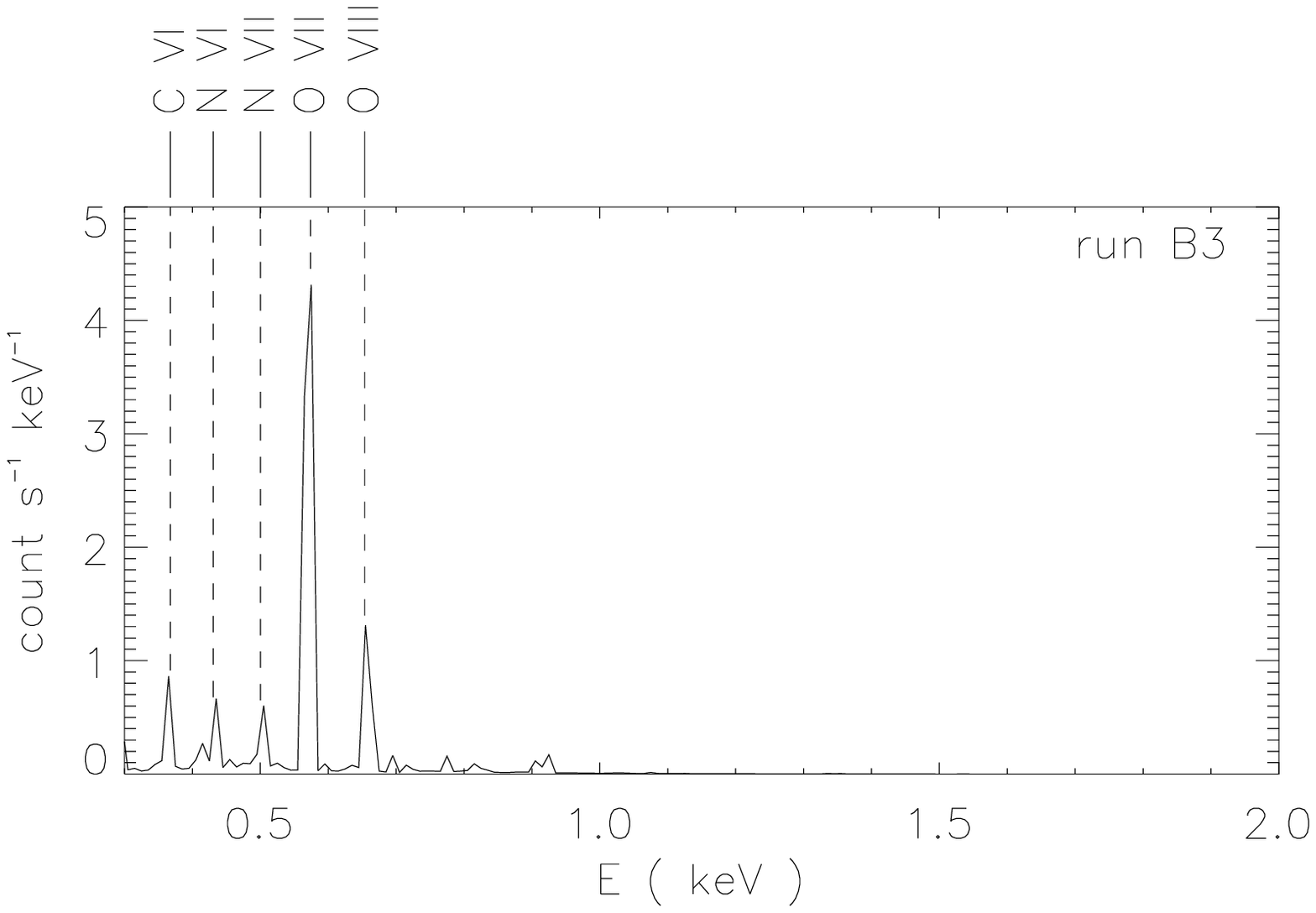}{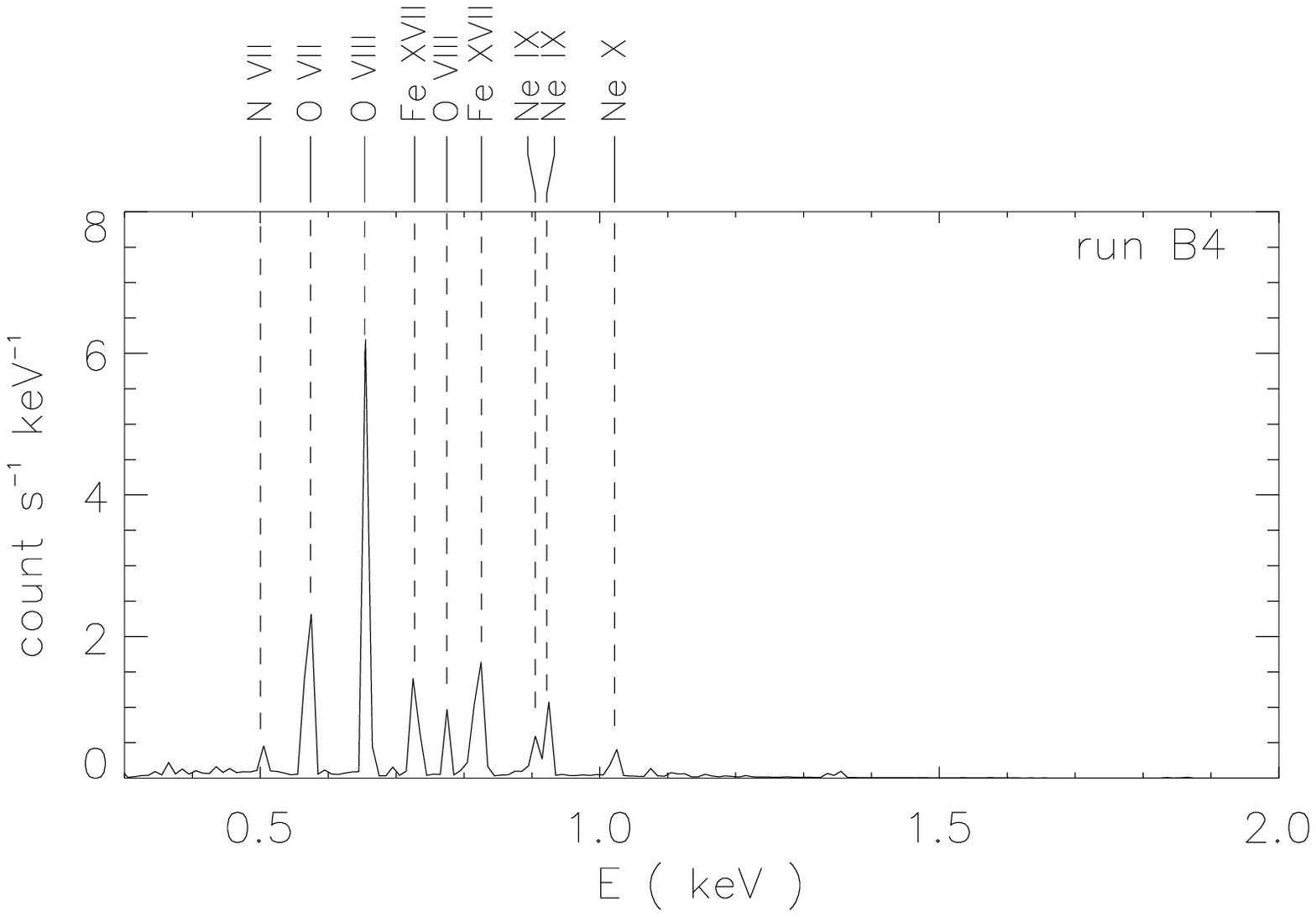}\\
\plottwo{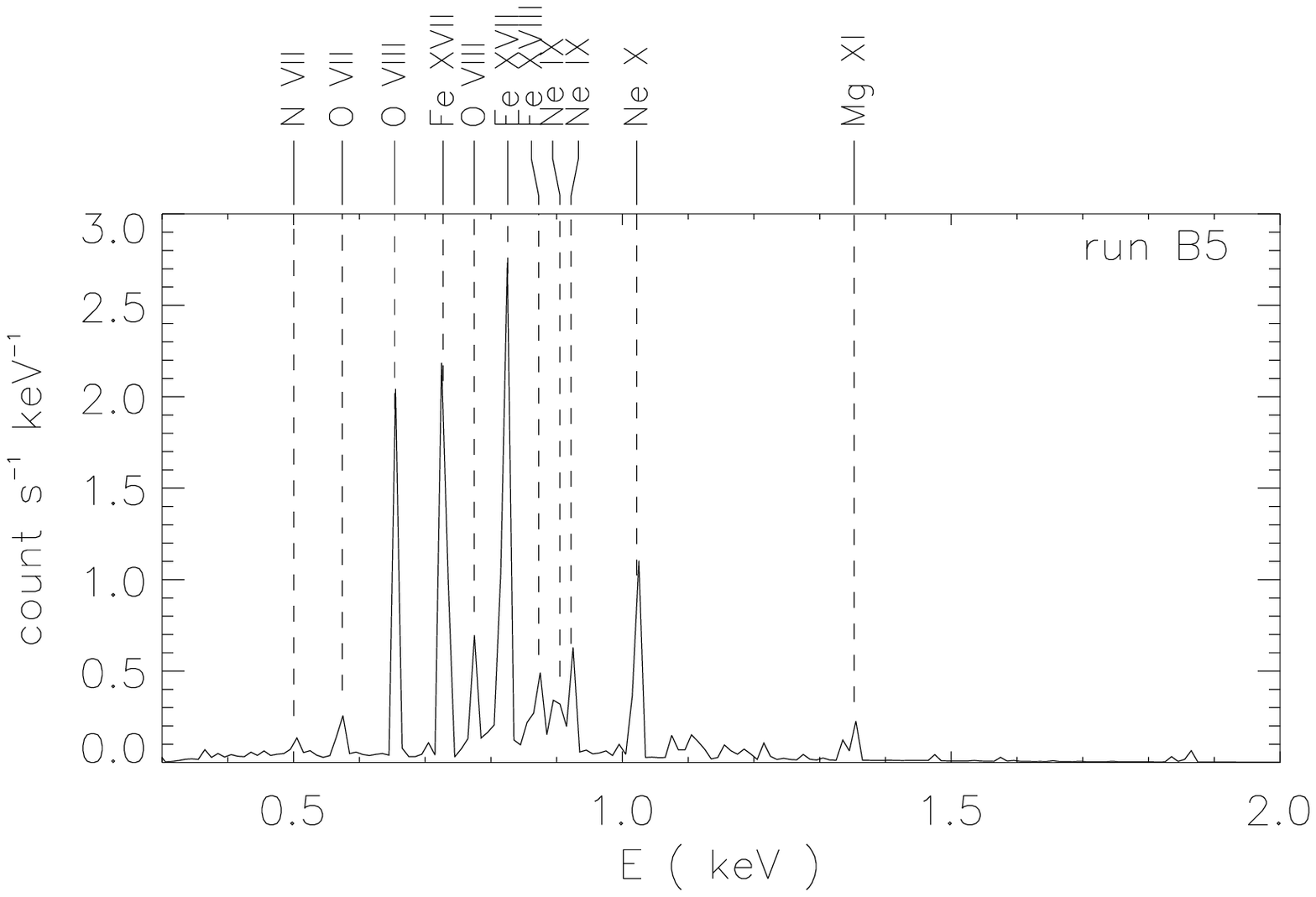}{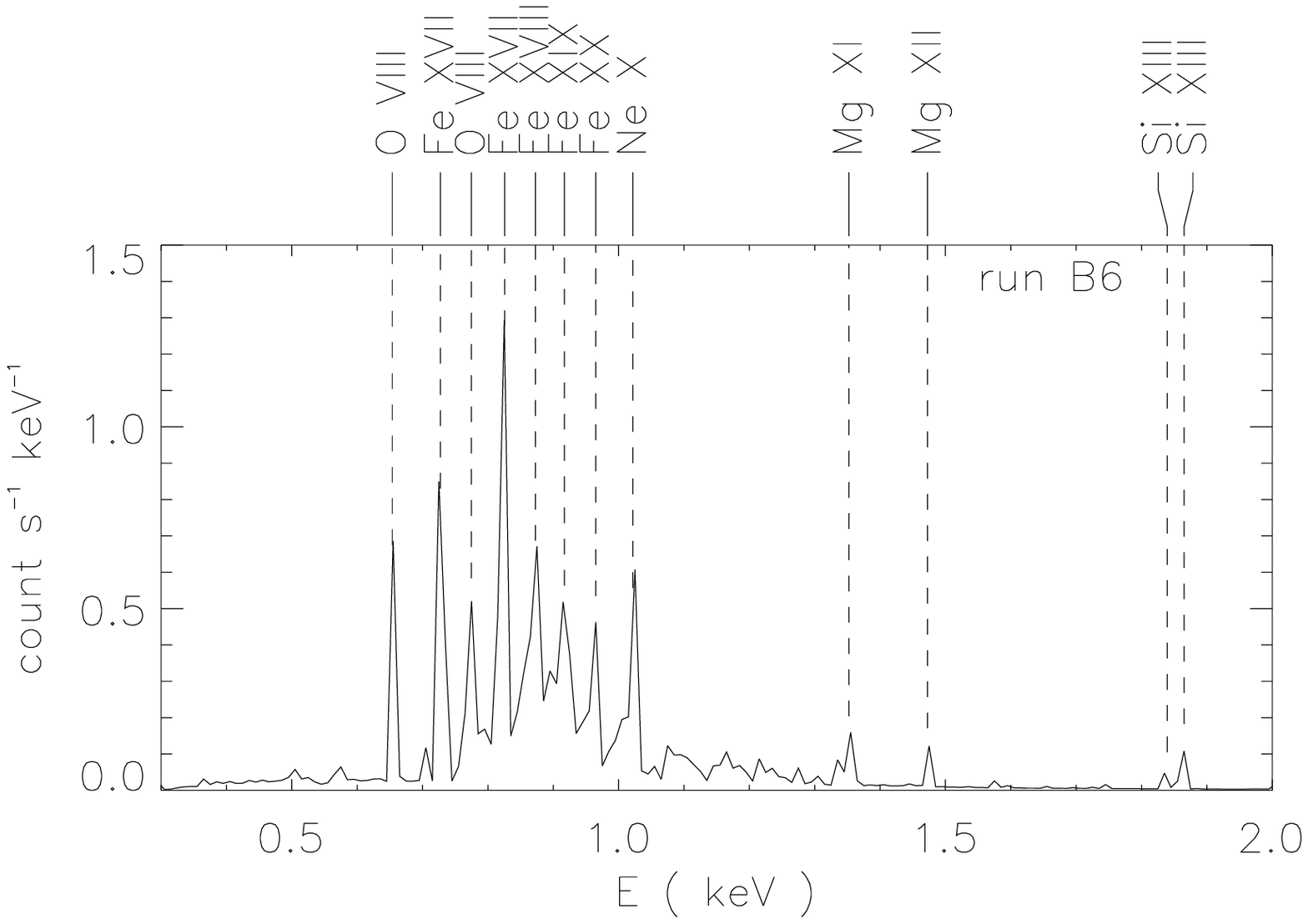}\\
\plottwo{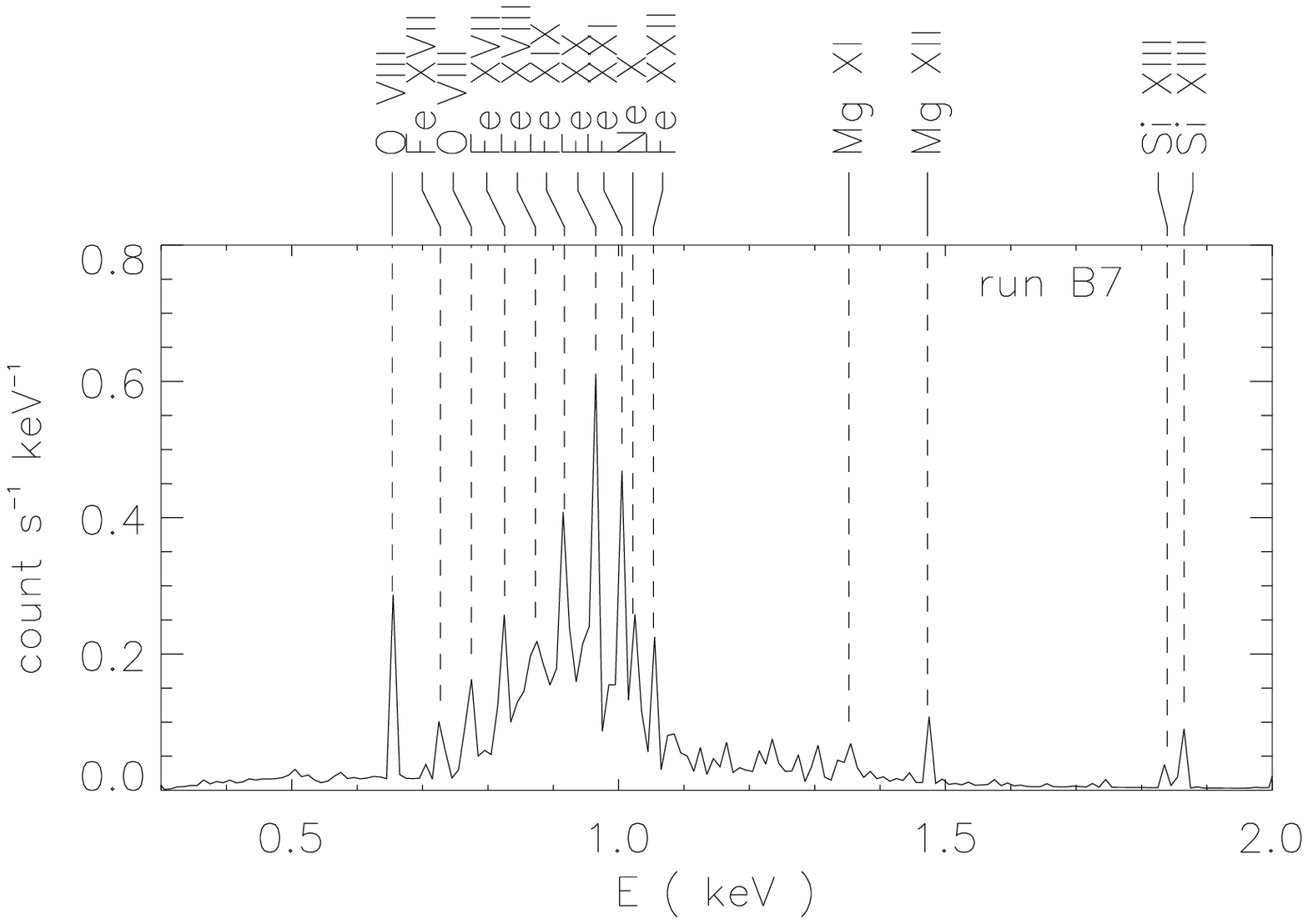}{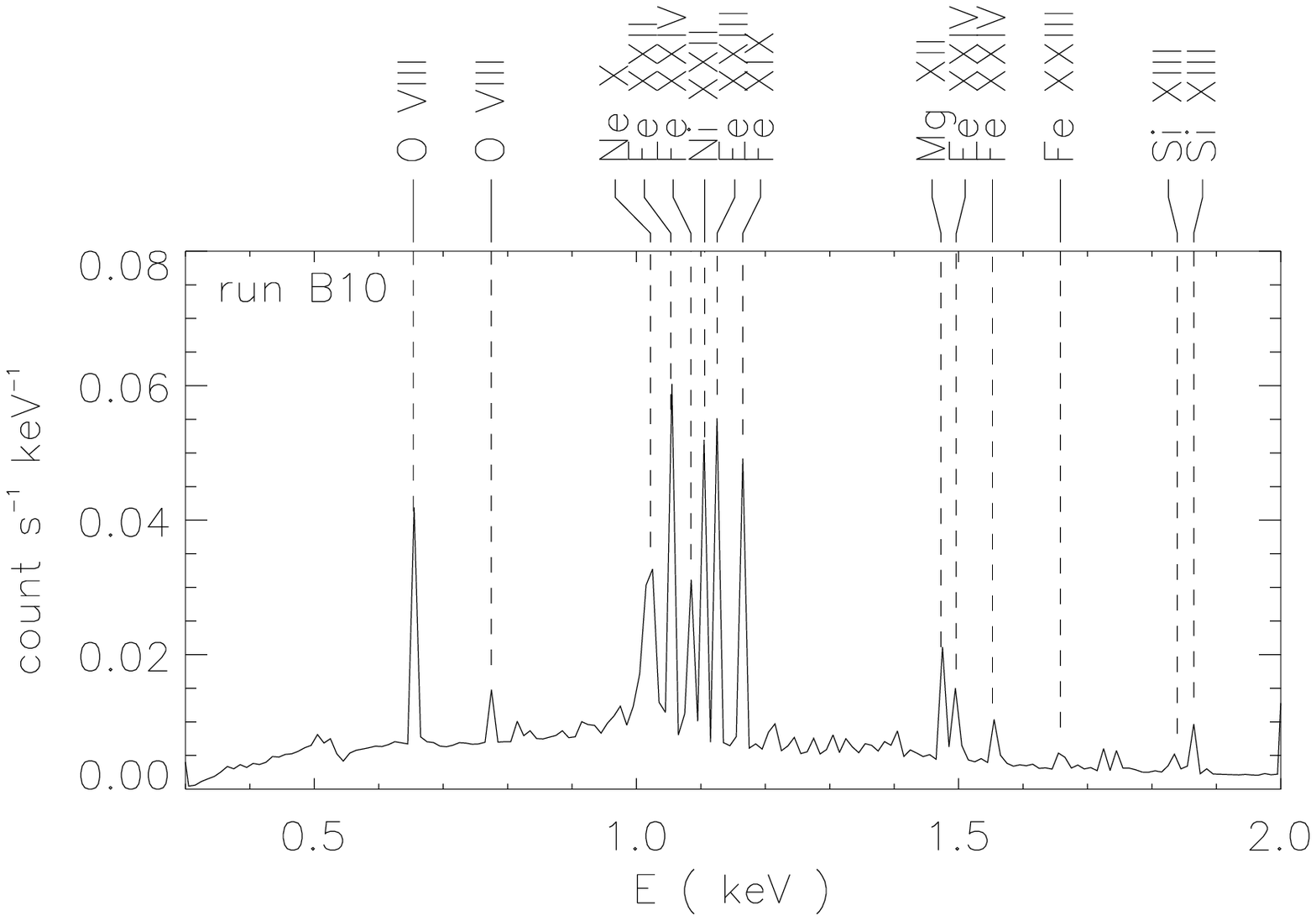}
\caption{X-ray spectra for the one-dimensional runs B3 -- B10; the model 
spectra have been convolved with the effective area of the {\em ACIS} S 
instrument (Fig. \ref{area_ACIS}); we assume a distance of 1 kpc; our spectra 
are presented with a resolution of 0.01 keV}
\label{X_spec_runs_1D}
\end{figure*}

As the temperature of the hot bubble is only a function of the fast wind
velocity, the spectra from, e.g., the models A5, B5 and C5, look almost 
identical. The differences between each set of models are in the absolute 
values of the flux, not the relative fluxes of different lines in the spectra. 

Within each set, as the value of $v_{\rm f}$ changes, the spectra also
change dramatically (Fig. \ref{X_spec_runs_1D}). At moderate velocities of the 
fast wind of 300 km s$^{-1}$, only a few lines with energies up to 1 keV are 
present. Increasing the fast wind velocity (and with it the temperature of the 
hot bubble) leads to the emergence of lines at higher energies, so that at and
above velocities of 600 km s$^{-1}$ one can see a peak of blended lines in the 
0.8 -- 1.1 keV range with an increasingly dominant contribution from iron 
lines. In the model B10, the continuum contributes to an energy 
range from 0.3 to above 2 keV. 

\section{Two-dimensional spherical models} \label{sec_res_2D}

In one-dimensional simulations, which we have discussed so far, instabilities 
and turbulence cannot occur. Instabilities may, however, be quite important in 
explaining microstructure in PNs, such as the observations of globules in a 
few PNs \citep{HuF05}. Therefore, we also calculated a two-dimensional 
spherical model\footnote{i.e. a two-dimensional coordinate system and 
spherical initial conditions} with the parameters of run B5 to examine 
the presence of 
such instabilities and how they affect the structure of the hot bubble and the
dense shell and the resulting X-ray properties.

\begin{figure}
\plotone{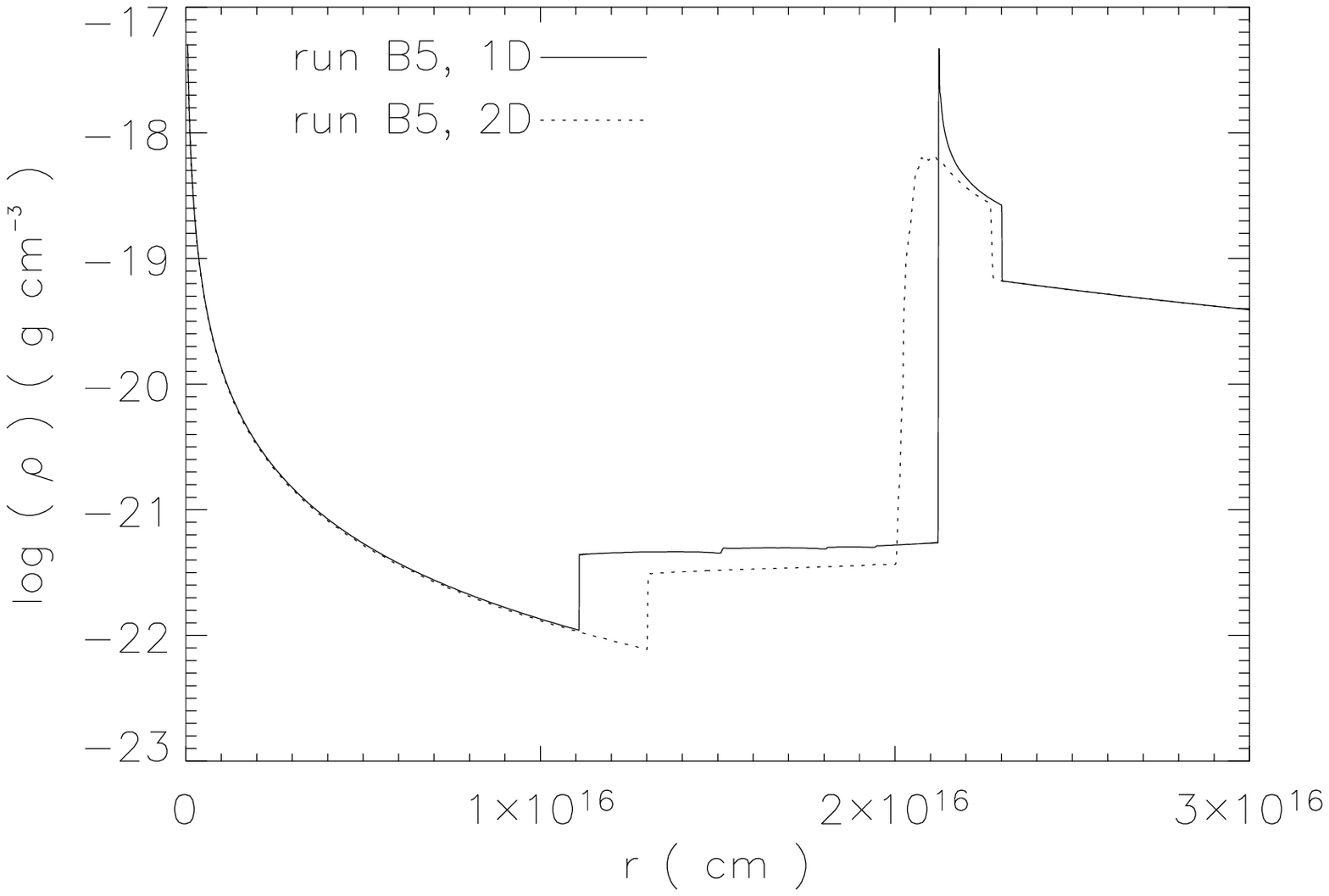}
\plotone{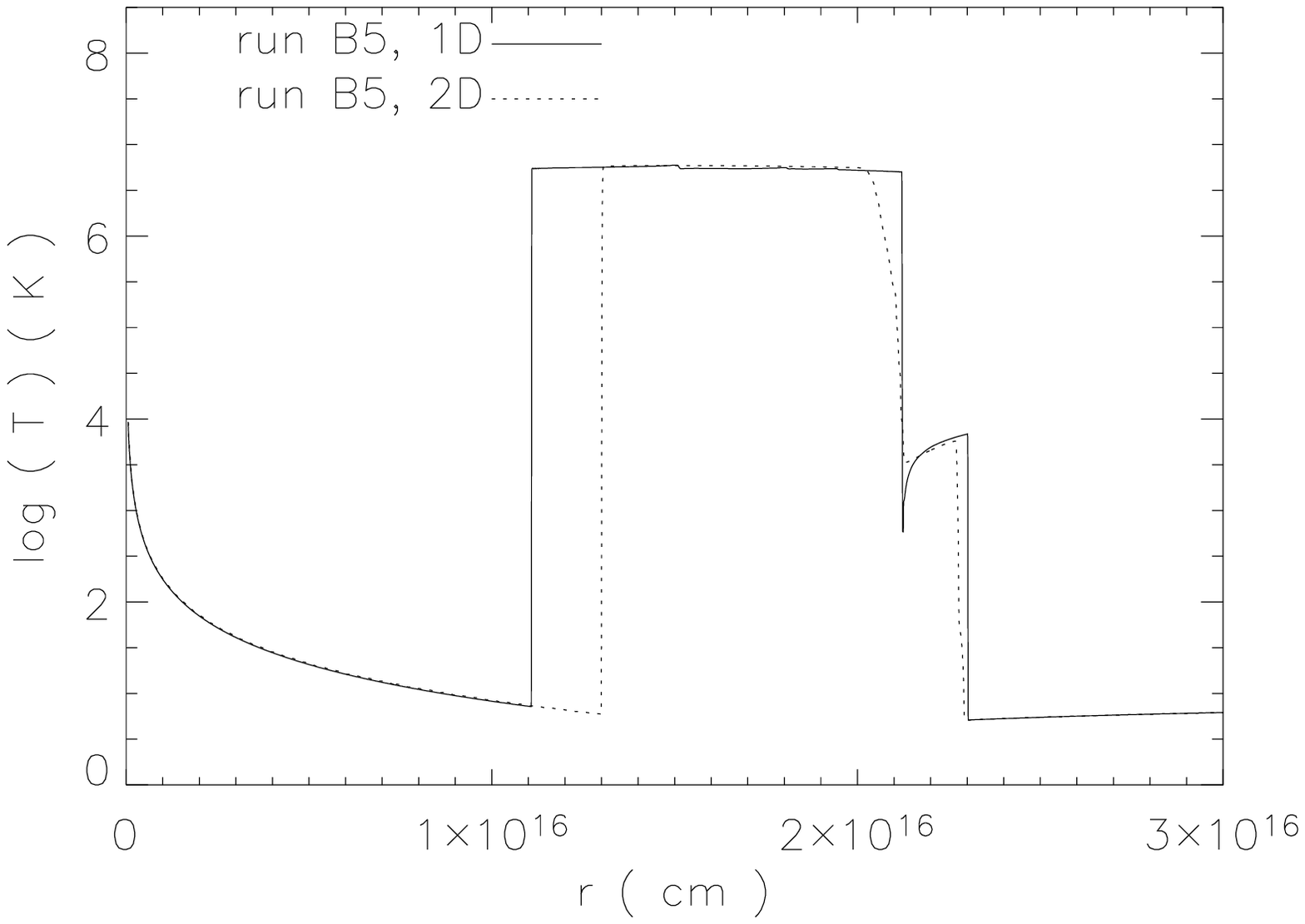}
\caption{Plots of the density and temperature as a function of radius at an 
age of 288 years 
of the one-dimensional (solid) and two-dimensional (dotted) run B5}
\label{comp_1D_2D_struc}
\end{figure}

The velocity of the shell is similar to the result of the one-dimensional 
simulation, but somewhat lower -- the value is 22.4 km s$^{-1}$ compared to 
about 27 km s$^{-1}$. As the expansion in the energy-driven phase is mainly 
due to the pressure in the hot bubble, the reduction of the shell velocity 
indicates that the pressure is reduced in the two-dimensional run.
The pressure in the one-dimensional run is $2.2 \times 10^{-7}$ dyn and $1.47 
\times 10^{-7}$ dyn in the two-dimensional run. The latter is reduced by about 
one third, which leads to a reduction of the shell velocity of 20 \%, as 
$v_{\rm shell} \sim \sqrt{p}$, i.e. a reduction from 27 to 22 km s$^{-1}$ as 
seen in the data. In Fig. \ref{comp_1D_2D_struc}, we show plots of the 
temperature and density as a function of radius at an age of 288 years of the 
one-dimensional run and an azimuthal average of the two-dimensional run -- as 
the temperature is identical, but the density is lower, the pressure is lower 
in the latter. The reduction of the pressure also produces a smaller extent of 
the hot bubble (i.e., the radial distance from the reverse shock to the 
contact discontinuity).

\begin{figure}
\plotone{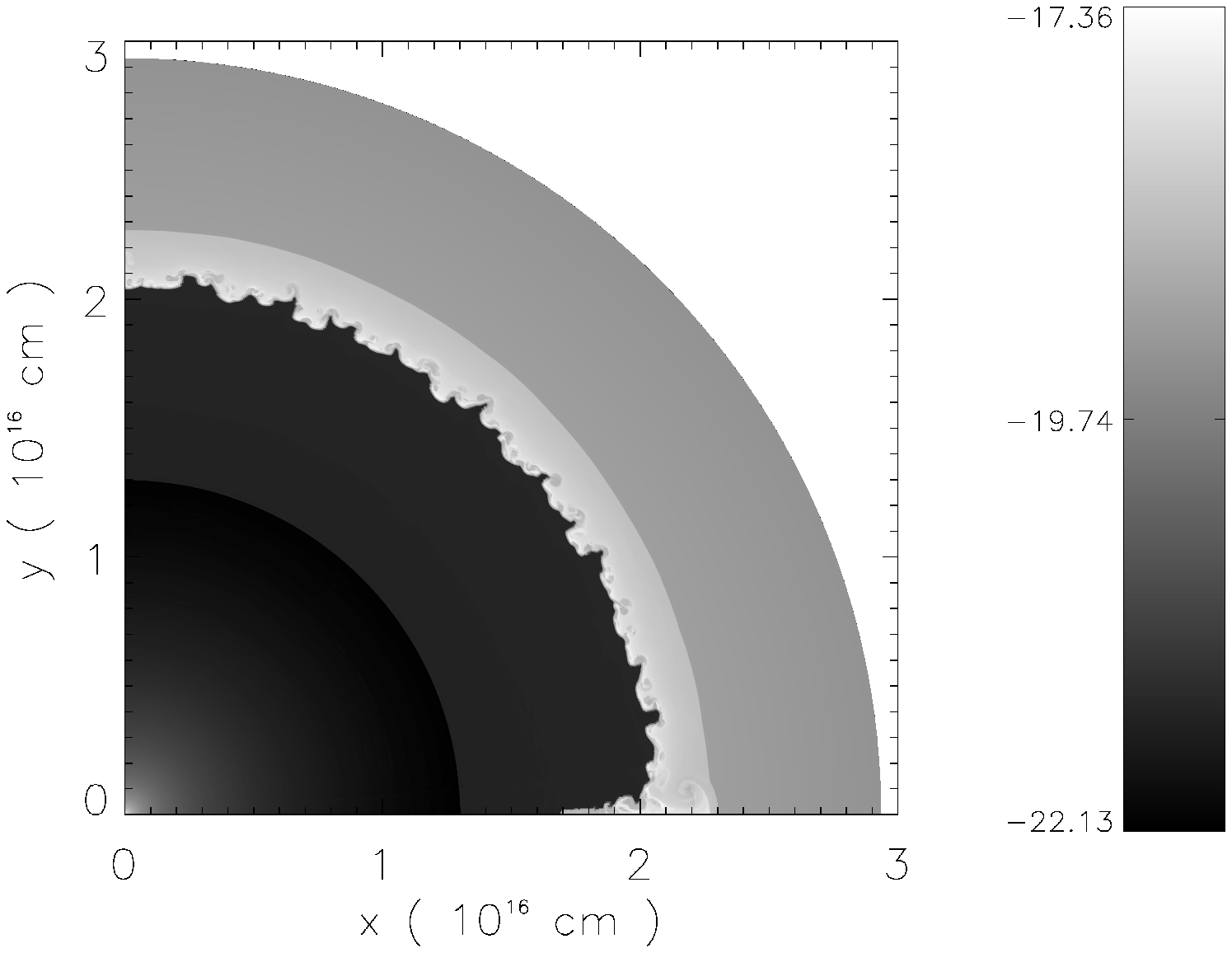}
\plotone{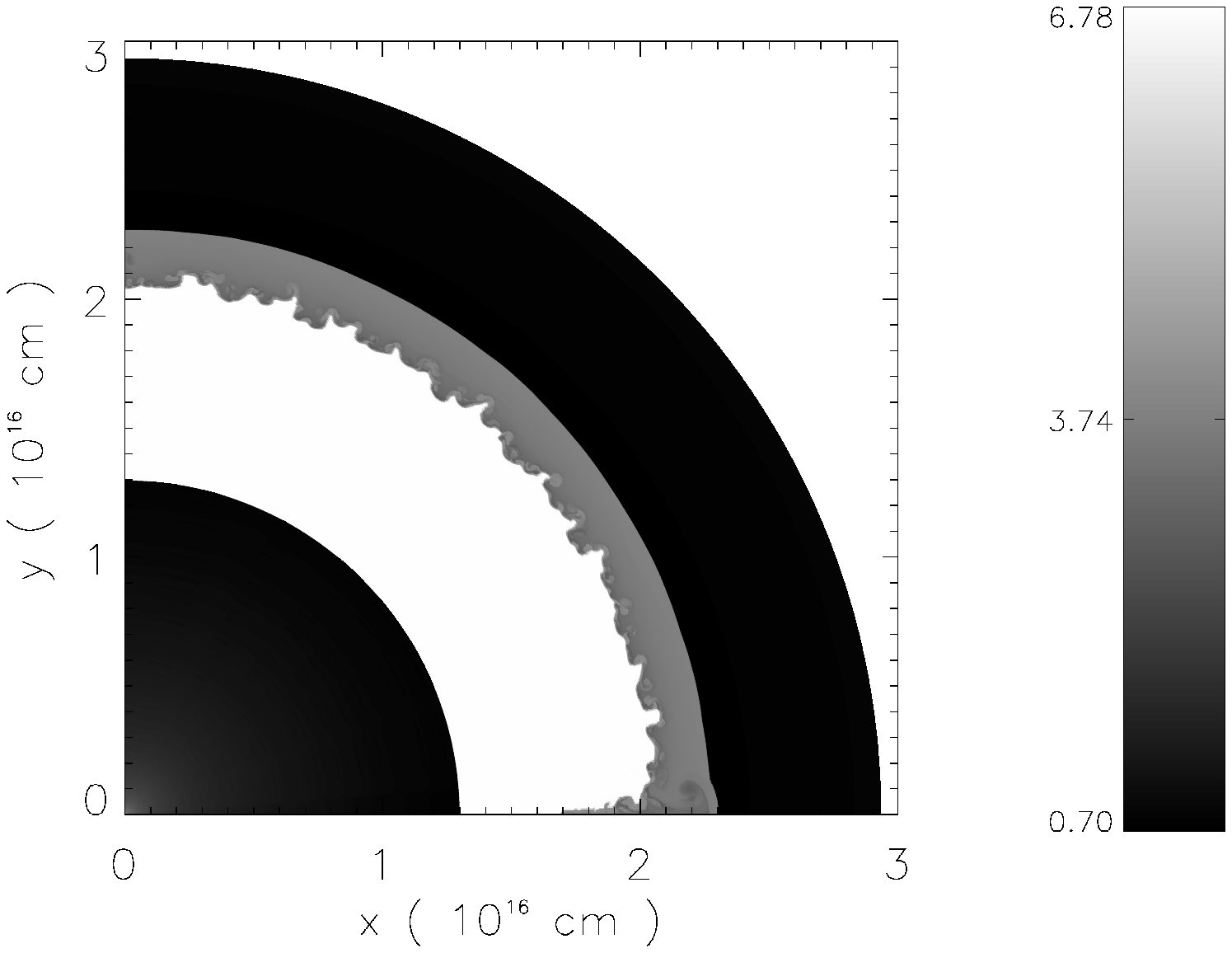}
\caption{Contour plots of the logarithm of density (top) and temperature 
(bottom) at an age of 288 years of the two-dimensional run B5}
\label{2D_den_T}
\end{figure}

The transition from a one-dimensional to a two-dimensional model 
introduces instabilities (Rayleigh-Taylor-instabilities), as shown in 
Fig. \ref{2D_den_T}. Due to these instabilities, some fraction of the total 
energy is redirected into the kinetic energy of non-radial motion. This 
fraction is typically a few percent and can be high as up to 40 \% locally in 
the transition layer between the hot bubble and the dense shell. Therefore 
less energy is available for the kinetic energy of radial motion of the dense 
shell and the internal energy of the hot bubble. Another effect 
triggered by instabilities is the mixing of material in the hot bubble and the
dense shell near the contact discontinuity which leads to a lowered 
temperature there. The consequent decrease in cooling time thus reduces the 
pressure and hence the extent of the hot bubble.

\begin{figure}
\plotone{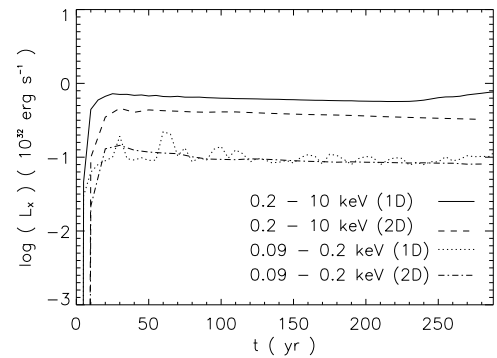}
\caption{X-ray luminosity $L_{\rm x,ACIS}$ in the energy bins 
between 0.09 -- 0.2 keV and 0.2 -- 10 keV as a function of the evolution time 
in the one-dimensional and two-dimensional runs B5}
\label{1D_2D_Lvst}
\end{figure}

The reduced extent of the hot bubble results in a reduced luminosity 
$L_{\rm x,ACIS}$ (by a factor of about 1.6, Fig. \ref{1D_2D_Lvst}). The 
luminosity in the medium energy bin (0.09--0.2 keV) remains almost unchanged. 
In the low energy bin (0.01--0.09 keV) the reduction of the extent of the hot 
bubble is compensated by the increased emission of lower temperature plasma 
resulting from turbulent mixing due to the instabilities, but we cannot 
quantify these effects accurately, because most of the energy in this energy 
bin is emitted by only a few grid cells in our models (see \S 
\ref{sec_res_1D_xray_lum}).

\section{Comparison with observations of apparently round PNs} \label{sec_appl}

In this section, we apply our one-dimensional models to two PNs which show 
apparently round morphology. Rather than computing new 2D runs, we used 1D 
runs in our modeling of BD$+$30$^\circ$3639 and NGC~40, since other 
uncertainties, as, e.g., those due to uncertain abundances, are much larger 
than the differences between the 1D and 2D runs. We use measured values of the 
fast wind properties (velocity, mass outflow rate) estimated from optical 
lines and derive the corresponding properties of the slow wind (using eqns. 
\ref{eq_vcr} and \ref{eq_vcde}) to fit the observed expansion velocity of the 
shell. In these simulations, we do not assume 
$\dot M_{\rm f}\,v_{\rm f} = \dot M_{\rm s}\,v_{\rm s}$.
Then we choose the evolutionary age in our simulation, where the size 
of the model dense shell is equal to the observed size of the PN, and compare 
the temperature of the hot bubble and the X-ray luminosity with values inferred
from available X-ray observations. Thus we constrain our models using four 
observed properties of PNs -- $v_{\rm exp}$, $r_{\rm shell}$, $T_{\rm x}$ and 
$L_{\rm x}$ -- whereas ASB06 take $T_{\rm x}$,  $L_{\rm x}$ and the dynamical 
age ($r_{\rm shell} /  v_{\rm exp}$) into 
account while comparing their models with the X-ray properties of PNs 
and ignore the measured values of the fast wind properties.

\subsection{BD$+$30$^\circ$3639}

BD$+$30$^\circ$3639 shows well-resolved, extended X-ray emission 
\citep{KSV00} which lies within the interior of the shell of ionized gas seen
in optical images. The extent of 
the shell in CHANDRA images is approximately 5''$\times$4''. Assuming a 
distance of 1 kpc \citep{KSV00}, the radius is therefore about 
$3.3\times 10^{16}$ cm. The fast wind velocity is about 700 km s$^{-1}$ 
\citep{LHJ96} and the mass loss rate of the fast wind is about $10^{-6}$ 
M$_\odot$ yr$^{-1}$ \citep{SoK03}. The average expansion velocity derived from 
[OIII] and [NII] emission lines is 25.5 km s$^{-1}$. \citet{KSV00} give a 
CHANDRA {\em ACIS} spectrum of BD$+$30$^\circ$3639 between 0.3 -- 1.7 keV and 
derive a total luminosity of $1.6 \times 10^{32}$ erg s$^{-1}$ after fitting 
the spectrum with a variable-abundance MEKAL model. The deduced 
emission-region temperature is $2.7 \times 10^6$ K. 

A new simulation D7 with the observed parameters was performed using 
$\dot M_{\rm s} = 7 \times 10^{-5}$ M$_\odot$ yr$^{-1}$. The observed 
size of the shell is reached at an age of 390 years. The model expansion 
velocity is 25.9 km s$^{-1}$, which is in good agreement with the 
observations\footnote{It is interesting to note that 
$\dot M_{\rm f}\,v_{\rm f} = \dot M_{\rm s}\,v_{\rm s}$ is satisfied in this
object.}. However, both the temperature of the hot bubble ($1.1 \times 10^7$ K)
and the luminosity ($9.2 \times 10^{32}$ erg s$^{-1}$ in the full range of 
{\em ACIS} of 0.2--10 keV, $7.8 \times 10^{32}$ erg s$^{-1}$ in the observed 
energy range) are too high in our model.

The model spectrum (Fig. \ref {lineid}, top) is different from the observed 
spectrum. In the observations, there is a strong, broad peak at about 0.4 keV,
which is not apparent in any of our model spectra. Furthermore there is a 
narrower, less prominent peak at about 0.9 keV in the observed spectrum,
however, our model spectrum shows a strong, broad peak at about 0.9 keV
(only visible in the models with a fast wind velocity above 600 km s$^{-1}$). 
The strongest lines in the region around the peak at 0.9 keV are iron lines 
(Fig. \ref{lineid}, top). This suggests that the iron abundance (and perhaps 
also those of other elements) in BD$+$30$^\circ$3639 are very different from 
the solar abundances \citep{AnG89}, which we have used in our models including 
run D7.

\subsubsection{The effects of uncertain abundances on X-ray emission 
properties} \label{sec_abund}

The abundance problem may be especially important for objects where the 
central star is of late [WC] type, as in BD$+$30$^\circ$3639 and NGC~40. In 
these objects, 
the abundances in the fast wind can be very different from those of the dense 
shell. The former, however, are very important for the X-ray properties, as the
hot bubble consists of material ejected in the fast wind. As mixing of material
from the dense shell into the hot bubble may occur, the abundances in the dense
shell also may be relevant for the X-ray properties. Such mixing has 
been inferred in the hydrogen-deficient PN Abell 30 by \citet{CCC97}.

An inspection of published values of abundances BD$+$30$^\circ$3639 shows that 
the abundances, both in the fast wind and the dense shell, are in general 
poorly determined. \citet{AlH95} estimated the following nebular abundances:
C and N are solar, O is roughly solar, Ne is depleted (although they only give
a lower limit) from a high spectral-resolution optical spectrum of
BD$+$30$^\circ$3639. \citet{BPW03} determined that C is enhanced by a
factor of 2, N and O are roughly solar, Ne is enhanced by a factor of 1.7,
using IR spectra and UV observations. 

The abundances in the fast wind are determined by \citet{LHJ96} by fitting 
optical spectra. They find an upper limit of the hydrogen abundance and that
helium, carbon and oxygen are very prominent as expected. \citet{ABH96} fit
ASCA observations with two sets of abundances: (i) a solar C and half-solar N
abundance, 0.2 solar for Ne, O and Fe are depleted; (ii) C is enhanced by a 
factor of 354, N and Ne are enhanced by a factor of 10, O is roughly solar and 
no iron. \citet{KSV00} concluded that C is enhanced, Ne is roughly solar and N 
and O are depleted, from a fit of its CHANDRA {\em ACIS} spectrum. 
\citet{MVK03} revisited the CHANDRA observations of \citet{KSV00} and showed 
that the fast wind abundances (the second set of abundances in Arnaud et al.
1996) can fit the observations much better than the nebular abundances of 
\citet{AlH95}. Furthermore they find that an enhanced Ne abundance relative to 
solar is required. \citet{GRA06}, re-investigating 
the same CHANDRA observations, also fit the spectrum with two sets of 
abundances: (i) C is enhanced by a factor of 3.7, Ne, N and O are depleted, 
only Si, Ca and Ni of the 7 heavier elements are solar, the others (Mg, S, Ar, 
Fe) are not existent; (ii) C is enhanced by a factor of 350, N and Ne are 
enhanced by a factor of 20, O is roughly solar, 7 heavier elements up to Ni 
are solar. In summary, the available data can be fit with a wide 
range of abundances, although there seem to be general agreement that carbon 
is enhanced and oxygen is either depleted or solar.
\begin{figure}
\plotone{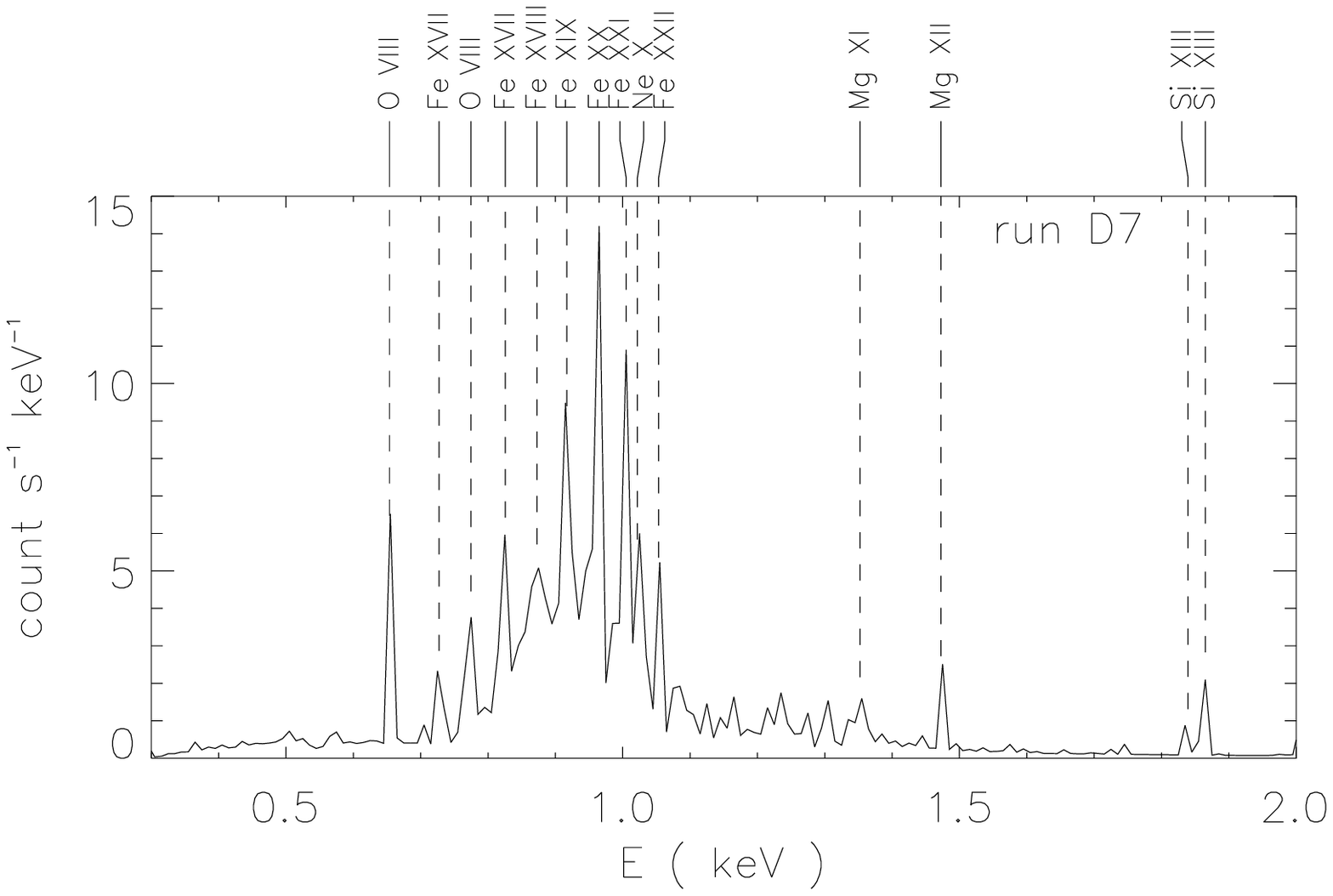}
\plotone{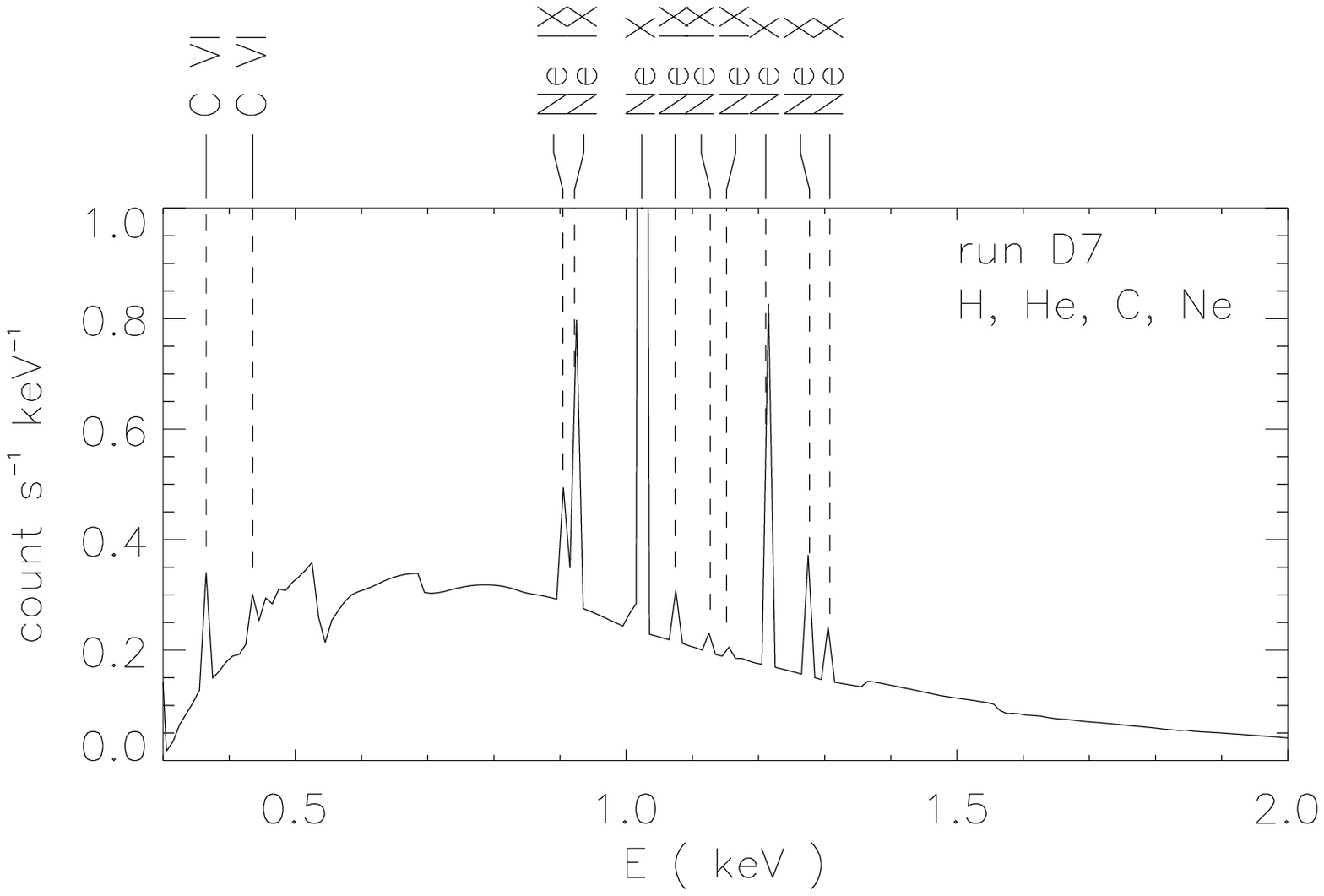}
\caption{X-ray spectrum for the one-dimensional runs D7 at an age of 390 years;
Top: 
calculation with the default set of elements and identification of the 
strongest lines; bottom: here only the elements H, He, C and Ne have been 
taken into account}
\label{lineid}
\end{figure}

In order to demonstrate the dramatic effect of changing elemental abundances on
X-ray emission properties, we have recalculated the spectrum from our model D7
using the elemental composition determined by \citet{KSV00} (i.e. only H, He, 
C and Ne, but with abundance ratios as in the Sun). The spectrum has the form
as shown in Fig. \ref{lineid} (bottom), which is closer to the observed 
shape. The total luminosity in the range of 0.3 -- 1.7 keV is then reduced to 
$1.75 \times 10^{32}$ erg s$^{-1}$, which is now very close to the observed 
value of $1.6 \times 10^{32}$ erg s$^{-1}$.

However, an important caveat to note here is that the cooling function 
depends on the abundances. For example, at temperatures $\gtrsim 10^6$ K the 
most important coolant is iron (see Fig. 18 in Sutherland \& Dopita 1993).
Hence an accurate calculation of X-ray properties as a function of 
abundances would require us to abandon the use of a cooling function,
replace it with a set of rate equations for the important species and 
calculate the resulting cooling for all transitions. This, however, would 
need a substantially increase in the computational effort, which is beyond 
the scope of this paper. 

Such an effort, however, might be warranted for modeling the high resolution 
spectrum of BD$+$30$^\circ$3639, recently taken using the {\em LETG} on 
CHANDRA, which shows abundance ratios of C/O $\sim 20$, N/O $\lesssim 1$, 
Ne/O $\sim 4$ and Fe/O $\lesssim 0.1$ \citep{KYH06}. The 
first step towards 
self-consistency of the cooling treatment would then be to modify the cooling 
function taking into account the effects of lower metallicities as in Fig. 13 
in \citet{SuD93}. 

\subsection{NGC~40}

Another example of a PN, which is detected in X-ray and is almost round, is 
NGC~40 \citep{MKD05}. Compared to BD$+$30$^\circ$3639, the fast wind velocity 
is significantly higher in this object 
(1000 km s$^{-1}$); the mass loss rate of the fast wind is 
$2.4 \times 10^{-6}$ M$_\odot$ yr$^{-1}$ \citep{LHJ96}. The observed
radius of the shell is 20,000 AU at an estimated distance of $\sim 1$ kpc. The 
average expansion velocity from [OIII] and [NII] emission lines is 27.5 km 
s$^{-1}$. The measured X-ray luminosity is $1.5 \times 10^{30}$ erg s$^{-1}$ 
in the range of 0.3--1 keV and the inferred temperature of the X-ray emitting 
gas is $1.5 \times 10^6$ K.

This object was modeled with a new run E10. We chose a mass loss rate of the 
slow wind of $5 \times 10^{-4}$ M$_\odot$ yr$^{-1}$ to adjust the expansion 
velocity of the dense shell in our model to the observed 
value\footnote{In this simulation, $\dot M_{\rm f}\,v_{\rm f}$ is not equal to 
$\dot M_{\rm s}\,v_{\rm s}$, but lower by a factor of 2.}. 
The expansion velocity is 27.03 km s$^{-1}$, which is very close to the 
observed one, as expected by our choice of the parameters. The observed size 
of the bubble was reached in our model after 3775 years, thus NGC~40 is much 
older than BD$+$30$^\circ$3639. 

We find from our model that the temperature of the hot bubble is 
$2.2\times10^{7}$ K and the luminosity is $2.4 \times 10^{33}$ erg s$^{-1}$ in 
the full range of {\em ACIS} (0.2--10 keV) and $7.6 \times 10^{32}$ erg 
s$^{-1}$ in the observed energy range. If, as in the case of 
BD$+$30$^\circ$3639, we restrict the composition to only H, He, C, N and O, 
the above luminosities are reduced to $1.5 \times 10^{33}$ erg s$^{-1}$ and
$6.4 \times 10^{32}$ erg s$^{-1}$, respectively. These values are still too 
high (by a factor of 400) to explain the observations. 

\subsection{The effects of different time histories of the fast wind} 
\label{sec_timehist}

The previous sections show that in our models of both objects, 
BD$+$30$^\circ$3639 and NGC~40, $T_{\rm x}$ and $L_{\rm x}$ are higher than 
the observed values. Noting that $T_{\rm x}$ is most sensitive to $v_{\rm f}$ 
(\S \ref{sec_res_1D_str} and eqn. \ref{eq_rh}), we investigate whether 
better fits to the data could be obtained by assuming that the value of 
$v_{\rm f}$ was lower in the past (thus implying that both objects are now
being observed in a phase where an even faster wind has emerged producing the
line profiles analyzed by \citet{LHJ96} for deriving $v_{\rm f}$).
We therefore compare two models, one with an ultra-fast 
wind of 1000 km s$^{-1}$ (run B10) and the second in which the fast wind has a 
velocity of 300 km s$^{-1}$ for the first 100 years followed by the ultra-fast 
wind (run B3+10). The mass loss rate of the slow wind was $7 \times 10^{-6}$ 
M$_\odot$ yr$^{-1}$. The momentum discharge 
$\dot M_{\rm f}\,v_{\rm f}=\dot M_{\rm s}\,v_{\rm s}$ is again kept constant. 

\begin{figure*}
\plottwo{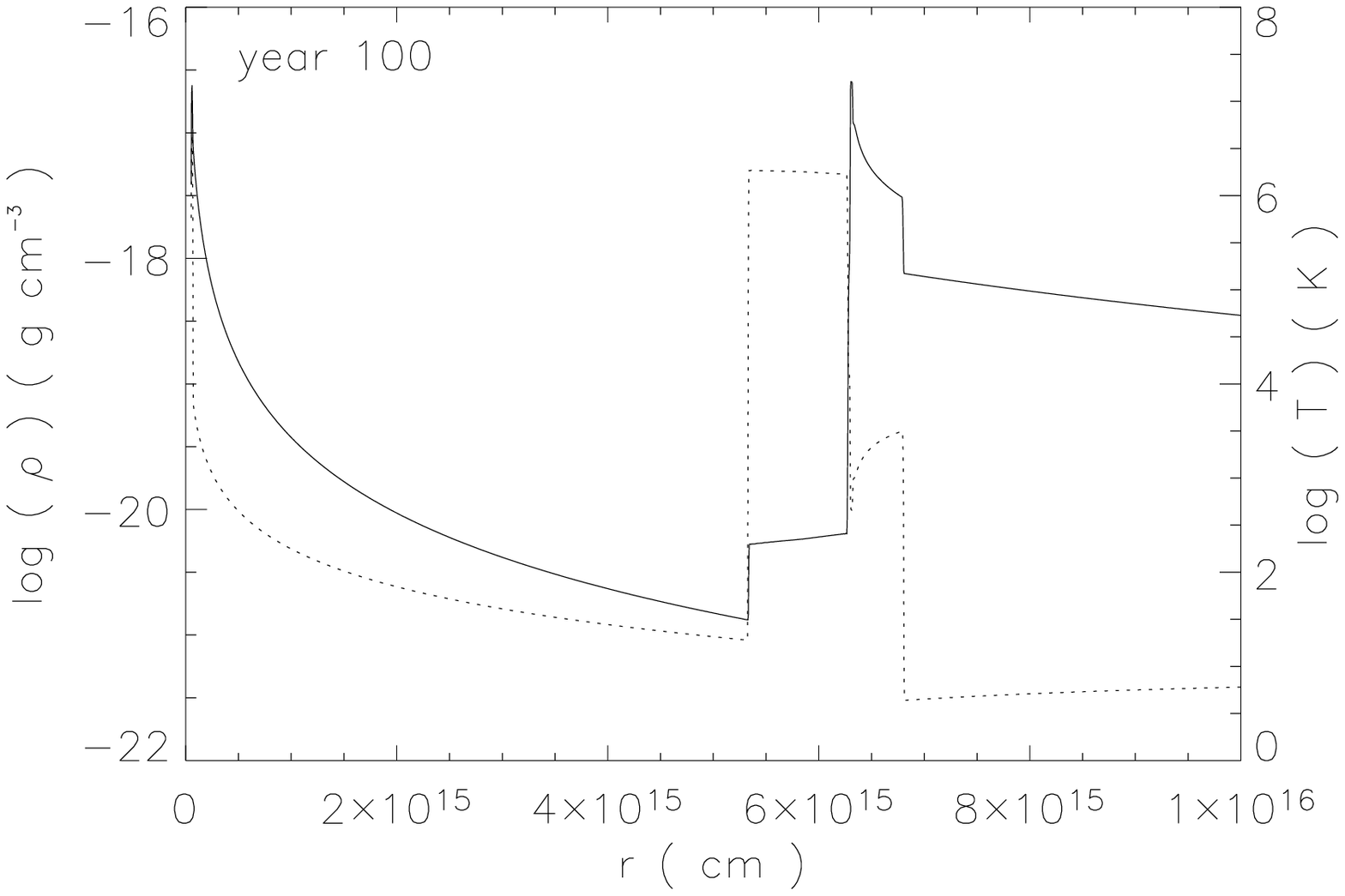}{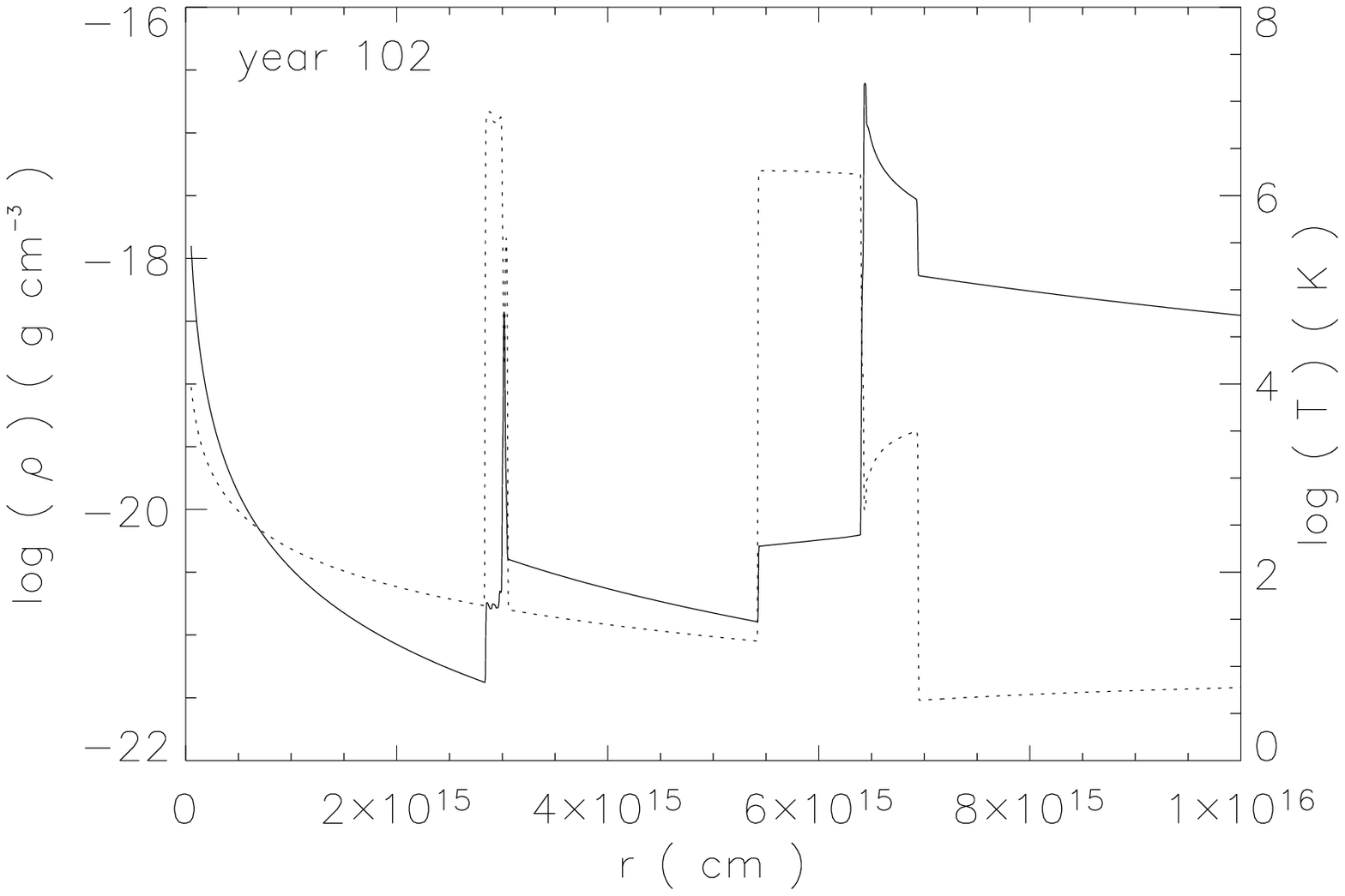}
\plottwo{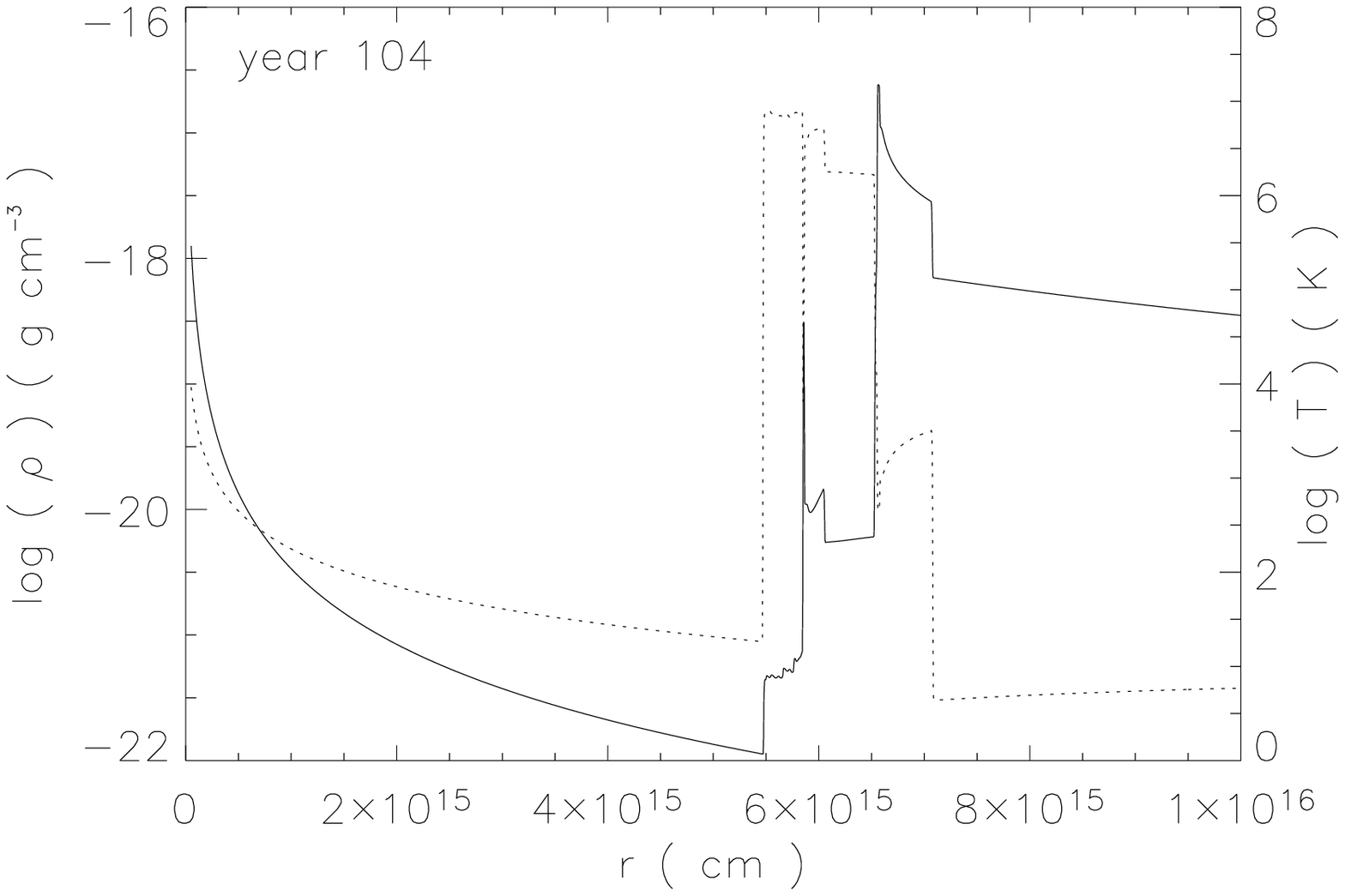}{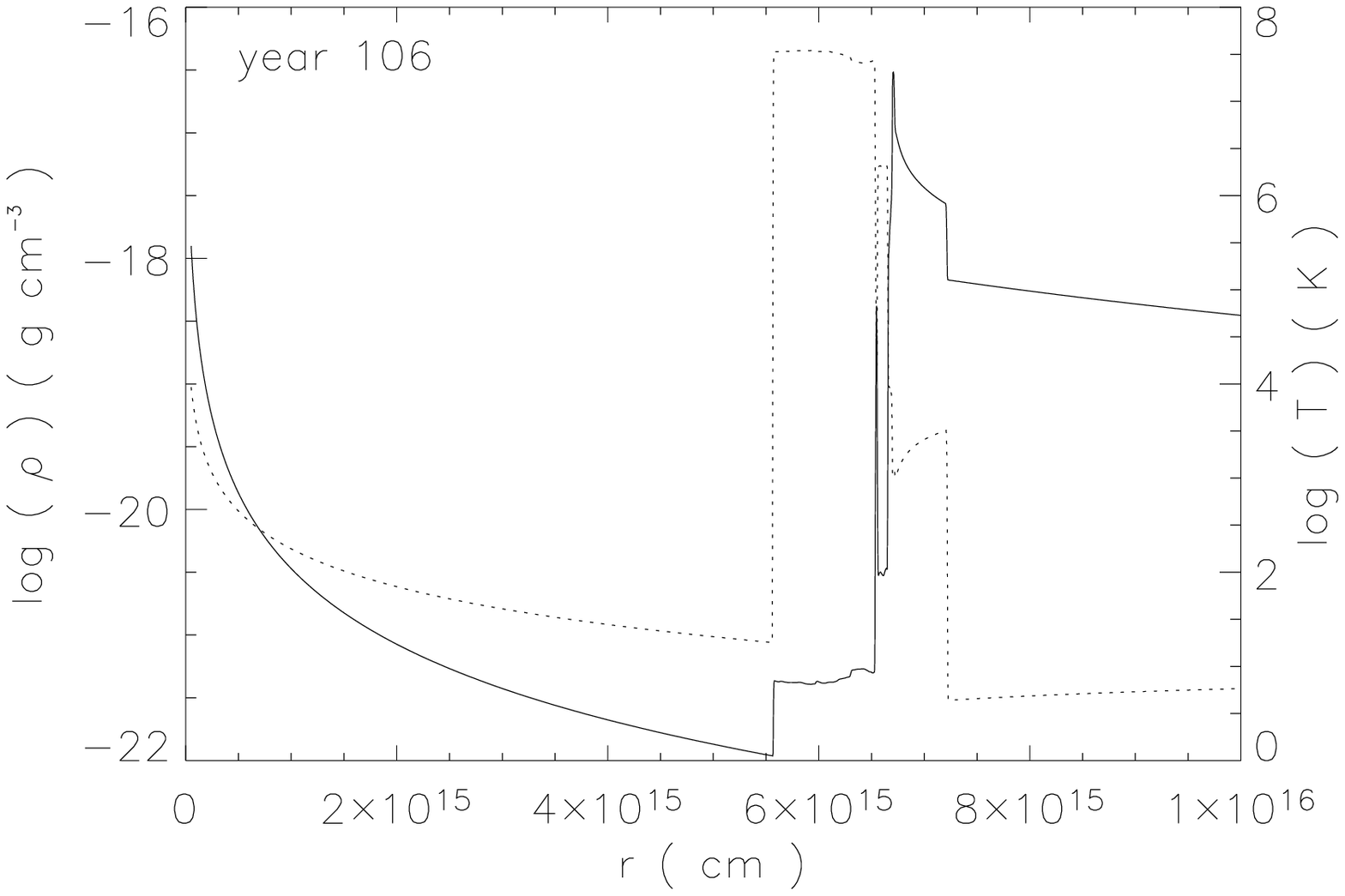}
\plottwo{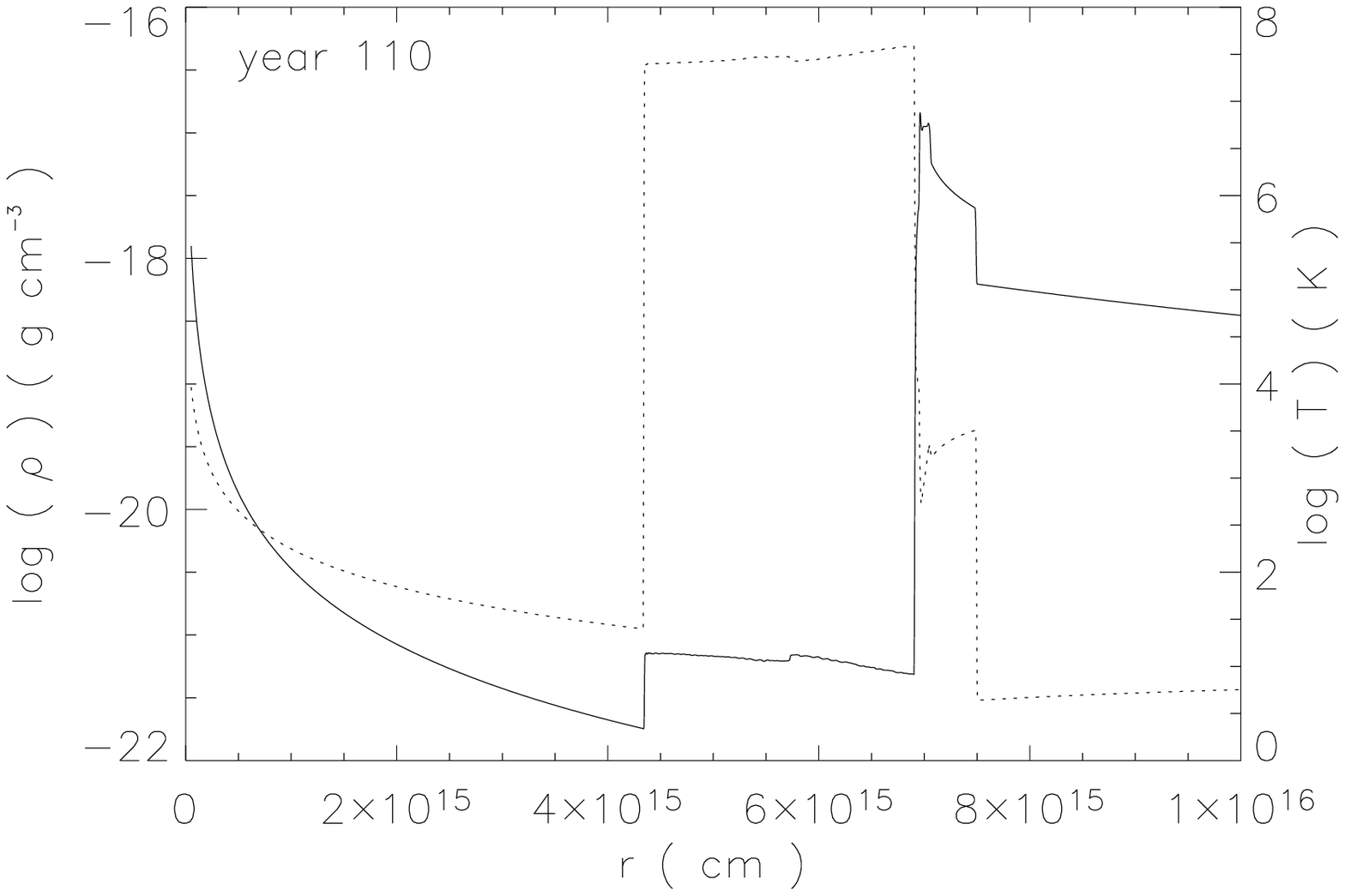}{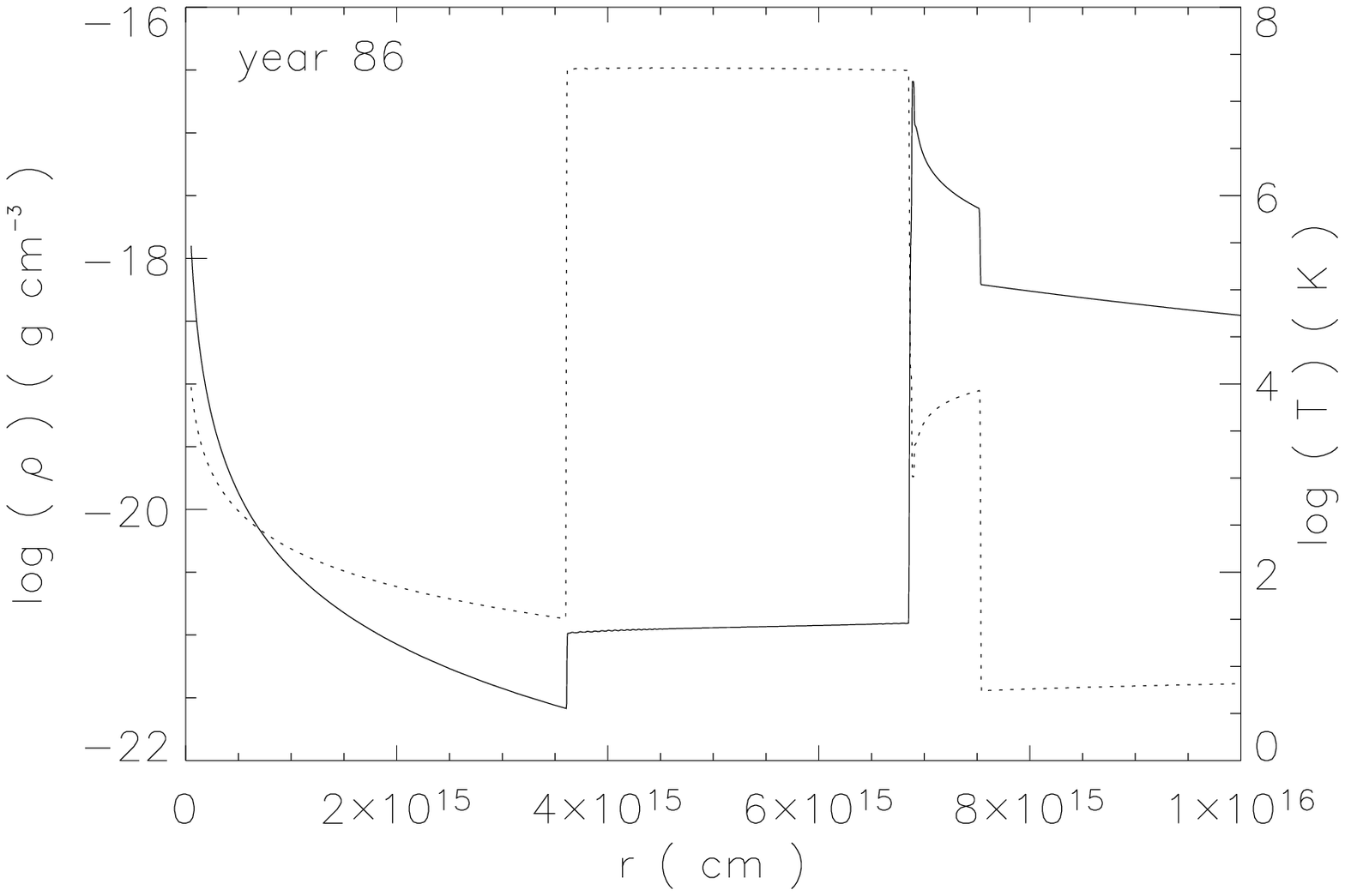}
\caption{Profiles of density (solid) and temperature (dotted) as a function of 
radius at ages of 100, 102, 104, 106 and 110 years for run B3+10 and for 
comparison at an age of 86 years for run B10 (bottom, right); the structure of 
the bubble is similar in both runs even directly after the merger}
\label{windwindwind}
\end{figure*}

We find that, in the second model, the contact discontinuity between 
the ultra-fast and the fast wind (CD$_2$) reaches the original contact 
discontinuity between the fast and the slow wind (CD$_1$)
in 7 years and, after a short period of complex interaction, 
produces a hot bubble with a structure very similar to that in the other model 
(Fig. \ref{windwindwind}).
The radial extent of the hot bubble is slightly smaller. The final expansion 
velocity of the shell is the same in both models (33 km s$^{-1}$). 

We extract the luminosities at early ages in the evolution of these models
(i.e., soon after emergence of the fast wind), when the size of the shell in 
each model has a value of 2000 AU.
The corresponding ages are 340 (in run B10) and 370 years (run B3+10). 
The simulated luminosities in the full range of {\em ACIS} are similar --
$8.9 \times 10^{30}$ erg s$^{-1}$ (run B10) and $9.3 \times 10^{30}$ erg 
s$^{-1}$ (run B3+10). The temperatures of the hot bubble are $2.1 \times 10^7$
K in both cases. Due to the slightly smaller extent of the hot bubble in run 
B3+10, the temperature gradient is slightly steeper. Although the spectra are 
very similar, a few small differences occur due to a larger quantity of cooler
gas in the run B3+10: enhanced emission is seen in the following lines at 0.22,
0.29, 0.36, 0.43 and 0.56 keV, which are lines of \ion{Si}{9}, \ion{C}{5}, 
\ion{C}{6}, \ion{N}{6} and \ion{O}{7}. These results show that, once CD$_2$ 
merges with CD$_1$, neither the expansion velocity 
of the shell nor the total X-ray luminosity or spectrum distinguish between
the very different time histories of the fast wind.

Observable differences are only present in the short period of time between the
emergence of the ultra-fast wind and the merger of CD$_2$ with CD$_1$.
We now have to check if it is likely that a specific PN is 
observed in this period. By using the radius of the dense shell 
at the time of emergence of the ultra-fast wind\footnote{since the velocity of 
CD$_1$ is very small 
compared to that of CD$_2$} and noting that CD$_2$ must 
expand faster than the fast wind, we can estimate 
a maximum time until the merger $t_{\rm coll}$. Thus for example, in our 
simulation B3+10, where the radius of CD$_1$ is 
about $6.3 \times 10^{15}$ cm after 100 years (Fig. \ref{windwindwind}),
\begin{equation}
t_{\rm coll} = \frac{6.3 \times 10^{15}\,\textrm{cm}}{300\,\textrm{km s}^{-1}} 
= 6.65\,\textrm{years}
\end{equation}
We derive $t_{\rm coll}$ for BD$+$30$^\circ$3639 and NGC~40, using the 
observed sizes of the two objects ($3.3\times10^{16}$ cm for 
BD$+$30$^\circ$3639 and $3\times10^{17}$ cm for NGC~40), and 300 km s$^{-1}$ 
for the minimum velocity of CD$_2$ since the temperature
of the hot bubble is better fitted by such a fast wind velocity (eqn. 
\ref{eq_rh}).
We find values of $t_{\rm coll}$ of about 36 for BD$+$30$^\circ$3639 and 
320 years for NGC~40. Since both values are a small fraction (one tenth) of 
the total age of the bubble in each object, it is very unlikely that 
the observations of \citet{LHJ96} caught these two objects in the special 
period after the emergence of the ultra-fast wind and before the merger of 
CD$_2$ with CD$_1$.

In summary we conclude that the utilization of a smaller value of $v_{\rm f}$ 
in the past will not help in explaining the low values of $L_{\rm x}$ and 
$T_{\rm x}$ observed in objects like BD$+$30$^\circ$3639 and NGC~40.

\subsection{The importance of heat conduction and its 
inhibition by magnetic fields}

Heat conduction across the contact discontinuity results in a cooler and 
denser hot bubble. \citet{ZhP96} give analytical estimates for the importance 
of heat conduction, calculating the ratio of the thermal conductivity time and 
the evolution time, claiming that heat conduction is important 
right from the beginning of the fast wind-slow wind interaction. However, they
ignore the effects of radiative cooling, which are considered by \citet{Sok94}.
By comparing the energy loss due to radiation and due to heat 
conduction from the hot bubble into the dense shell, \citet{Sok94} finds
that radiative losses become significant compared to heat conduction after a 
certain timescale $t_{\rm cool}$ (his eqn. 6),
when the conduction front thickness (which increases with time) reaches a 
certain value. He derives this timescale by using the growth rate 
of the conduction front from a model by \citet{Bal86}. However, both 
\citet{Sok94} and \citet{Bal86} assume that the pressure of the hot gas is 
constant with time. Since the pressure of the hot bubble is expected to vary 
strongly with time in PNs ($\sim t^{-2}$), the formulation of \citet{Sok94} 
may not be valid.

The heat conduction front goes through three different stages. In the first 
stage, mass from the cold shell is evaporated into the hot bubble. This results
in the hot bubble mass being dominated by the evaporating gas from the dense 
shell \citep{ZhP96}. Thus in a PN in this stage the X-ray spectrum should 
reflect the abundances of the dense shell and not those of the shocked fast 
wind. This is also true for the second, quasi-static stage. In the third stage,
i.e. ``condensation'', the region of intermediate temperature expands into the 
region of the initially hot gas \citep{BBF90}. In the last two stages 
radiative losses dominate heat conduction. \citet{SoK03} claim that in 
BD$+$30$^\circ$3639 the inferred depleted abundance of oxygen, based on the 
modeling of the CHANDRA {\em ACIS} spectrum by \citet{KSV00}, suggests 
that the X-ray emitting conduction front is in one of the first two phases, 
i.e. located within the dense shell, making the implicit assumption that 
oxygen is depleted in the dense shell. However, we do not think that their 
claim is justified as the oxygen abundance seems to be roughly solar in the 
nebula, i.e. the dense shell (\S \ref{sec_abund}). The abundances as 
inferred from the new high-resolution X-ray spectra definitely show a very 
large overabundance of carbon (as expected in the fast wind from the [WC] 
central star of this object) in contrast to the carbon abundance of the dense 
shell (\S \ref{sec_abund}). If heat conduction is operating in this object,
it can not have evaporated a large amount of mass from the dense shell into 
the hot bubble. Alternatively, heat conduction is 
inhibited by magnetic fields in this object.

Magnetic fields, which are believed to play an important role in collimating 
the fast outflow, may reduce thermal heat conduction {\em across} field lines 
\citep[e.g.][]{Sok94}. Following \citet{BBF90}, the conductivity coefficient 
can be written
\begin{equation}
\kappa = \frac{C\,T^{5/2}}{1 + ( n\,\tan\,\theta_{\rm h} / n_{\rm h} )^2 }
\end{equation}
Using the expression for the gyroradius $\delta = \sqrt{2\,T\,
m_e}/ ( e\,B )$ with the electron mass $m_e$ and the magnetic field strength 
$B$
\citep{Spi62}, one can show that a very weak magnetic field of the order of 
$0.1\,\mu G$ is sufficient to reduce the conductivity \citep{Sok94}. 
\citet{ChL94} find in the models of magnetized stellar wind bubbles that the 
tangential magnetic field component in the shocked fast wind is larger than 
the radial component by several orders of magnitude, thus heat conduction is 
inhibited in radial direction.

Heat flux inhibition by electromagnetic instabilities can also occur
{\em along} the field lines, if $\beta = 8\,\pi\,p / B^2 \gg 1$ \citep{LeE92}, 
which gives a maximum magnetic field
\begin{eqnarray}
B &\ll& \sqrt{8\,\pi\,p} = \sqrt{8\,\pi\,n\,k\,T} = \nonumber \\
&=& 5.88\, \mu G\,\left( 
\frac{n}{100\,\textrm{cm}^{-3}} \right)^{1/2}\,\left( 
\frac{T}{10^6\,\textrm{K}} \right)^{1/2}
\end{eqnarray}
When the magnetic field is strong, the suppression is small \citep{PiE98}.

Several lines of evidence support the presence of magnetic fields in PNs 
and therefore the inhibition of heat conduction is not unlikely.
For example, magnetic fields in PPNs have been inferred from the presence of 
polarized OH and H$_2$O masers \citep[e.g.][]{VDI06} and have been detected in 
the central stars of PNs \citep{JWT05}. Magnetic fields are also considered to 
be the main agent for producing bipolarity in PNs \citep{GLF05}. 

\section{Comparison with analytical results} \label{sec_diff}

We compare our results now with the analytical results of ASB06. A basic 
advantage of an analytical approach is that it allows us to explore a large 
part of the parameter in a short time. However, the analytical approach 
requires certain assumptions which at best may be only partially valid. 
Numerical simulations like those which we have carried out in this paper, 
although much more time consuming, are thus necessary for testing the 
limitations of the analytical approach as well as making realistic models of 
real objects.

ASB06 calculated the structure of the hot bubble and the dense shell using 
self-similar solutions as in \citet{ChI83}. They computed the X-ray luminosity 
between 0.2 -- 10 keV (now called high energy bin) by taking into account only 
those gas parcels in which the cooling timescale (from the cooling 
function of Sutherland \& Dopita 1993) is larger than the evolution timescale 
of the flow.
They derived the X-ray emissivities from the {\em APEC} database \citep{SBL01} 
and integrated the emission measure over the region between the inner shock and
the contact discontinuity. Their results can now be compared with those from 
our simulations. 

We find a different temporal behavior of the luminosity in the high energy bin,
$L_{\rm x,ACIS}$, compared to the result of ASB06 -- whereas $L_{\rm x,ACIS}$
in our models increases slightly over the first few 100 years and 
decreases slowly ($\sim t^{-0.35}$) after that, in ASB06's models it  
decays with time as $\sim t^{-1}$. These differences probably arise because 
their cooling treatment is not fully self-consistent. They do not follow the 
evolution of the gas parcels with radiative cooling, as of course is done in 
our numerical simulations. Furthermore the temporal behavior of the luminosity 
in ASB06's models depends on the specific criteria for removing cooled regions 
(see their Fig. 9). 

The initial phase of low $L_{\rm x,ACIS}$ is of significantly smaller 
duration in our models as compared to ASB06 -- e.g. in model C3, it covers a 
range of only 100 years in our models, whereas in ASB06 it is about 700 years. 
This difference is relevant to the understanding of X-ray emission from PPNs 
where the ages of the nebulae are typically $\sim$ few $\times 100$ years.
The length of the phase of increasing luminosity is dependent on the parameters
of our models -- the lower the velocity and the higher the mass outflow rate 
of the fast wind, the longer this phase. Both trends are qualitatively 
consistent with the results of ASB06.

After the initial phase, a little step in the luminosity occurs (e.g. in 
run B3 at an age of about 230 years in Fig. \ref{Lvst_allruns}).
Shortly before this transition, oscillations of density, pressure and velocity 
occur in the hot bubble. \citet{FrM94} have also found oscillations in their 
numerical simulations and suggest that these occur when 
energies, e.g. the total and the internal energy, are almost equal. In our 
case, the internal energy is about 96 \% of the total energy. However, whether
these oscillations are only symptoms of, or the reason for, the transition, 
is unclear. Investigating the age at which the steps occur as a function of the
parameters of the fast wind, we find that 
$t_{\rm step} \sim \dot M_{\rm f}^{0.5}\,v_{\rm f}^{0.75}$. Although we can not
say as yet if the oscillations are a numerical artefact or a real physical 
effect, these scalings might help in eventually finding an answer.

The temporal behavior of $L_{\rm x,ACIS}$ in the models of ASB06 and our 
models can be easily understood as follows. In the self-similar models of 
ASB06, the radius of the contact discontinuity and that of the 
reverse shock are proportional to $t$, i.e. the volume of the hot bubble 
varies as $t^3$. The density decreases as $t^{-2}$, hence the emission measure 
$n^2\,V$ varies as $t^{-1}$. As the temperature in the hot bubble remains 
roughly constant, the emissivity $\Lambda$ of the hot bubble gas is 
also constant and therefore the total luminosity ($\Lambda\,n^2\,V$) varies as 
$t^{-1}$. In comparison, the empirical scaling laws in our runs
are $r_{\rm rs} \sim t^{0.98}$, $r_{\rm cd} \sim t^{1.08}$, 
$V = \left( r_{\rm cd}^3 - r_{\rm rs}^3 \right) \sim t^{3.35}$, 
$< \rho > \sim t^{-1.86}$ and $< \rho >^2\,V \sim t^{-0.37}$, resulting in 
$L_{\rm x,ACIS} \sim t^{-0.37}$. 
The scaling laws of density and volume of the hot bubble imply 
that the total mass of shocked fast wind material increases as about 
$\sim t^{1.5}$, whereas the mass input by the fast wind increases only 
as $\sim t$, since the mass outflow rate is kept constant. This difference in 
the time dependence reflects the evolution of the hot gas in the shocked fast 
wind under the influence of radiative cooling. In the beginning, only a small 
fraction of the mass injected by the fast wind is at high temperatures in the 
hot bubble, most of it lies near the contact discontinuity at low temperatures.
This is because, due to radiative cooling, large parts of the shocked fast 
wind collapse towards the contact discontinuity. With increasing age, the 
density decreases and this effect becomes weaker, progressively increasing the 
fraction of gas injected by the fast wind which remains at high temperatures 
after being shocked.
\begin{figure*}
\includegraphics[width=0.5\textwidth]{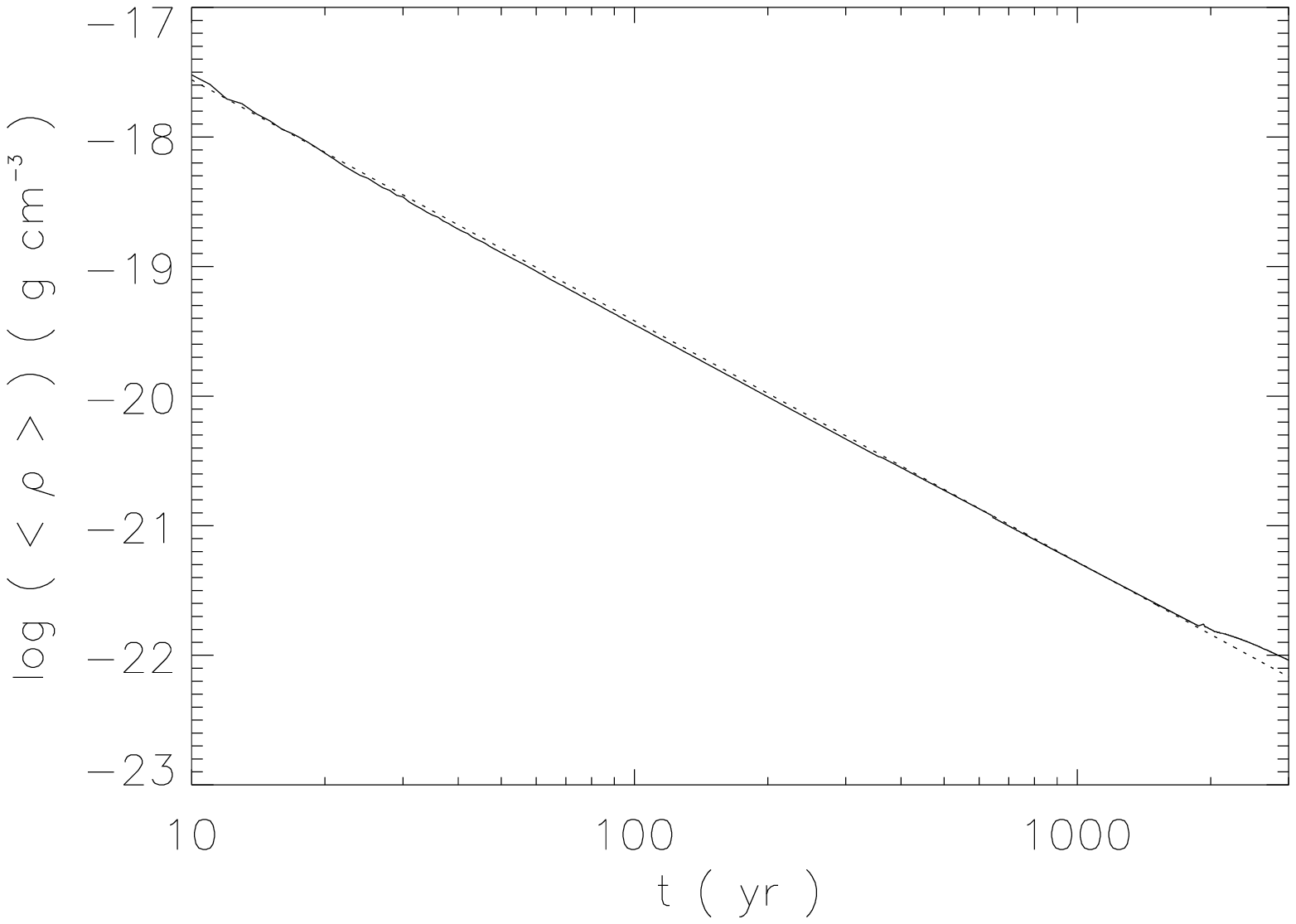}
\includegraphics[width=0.5\textwidth]{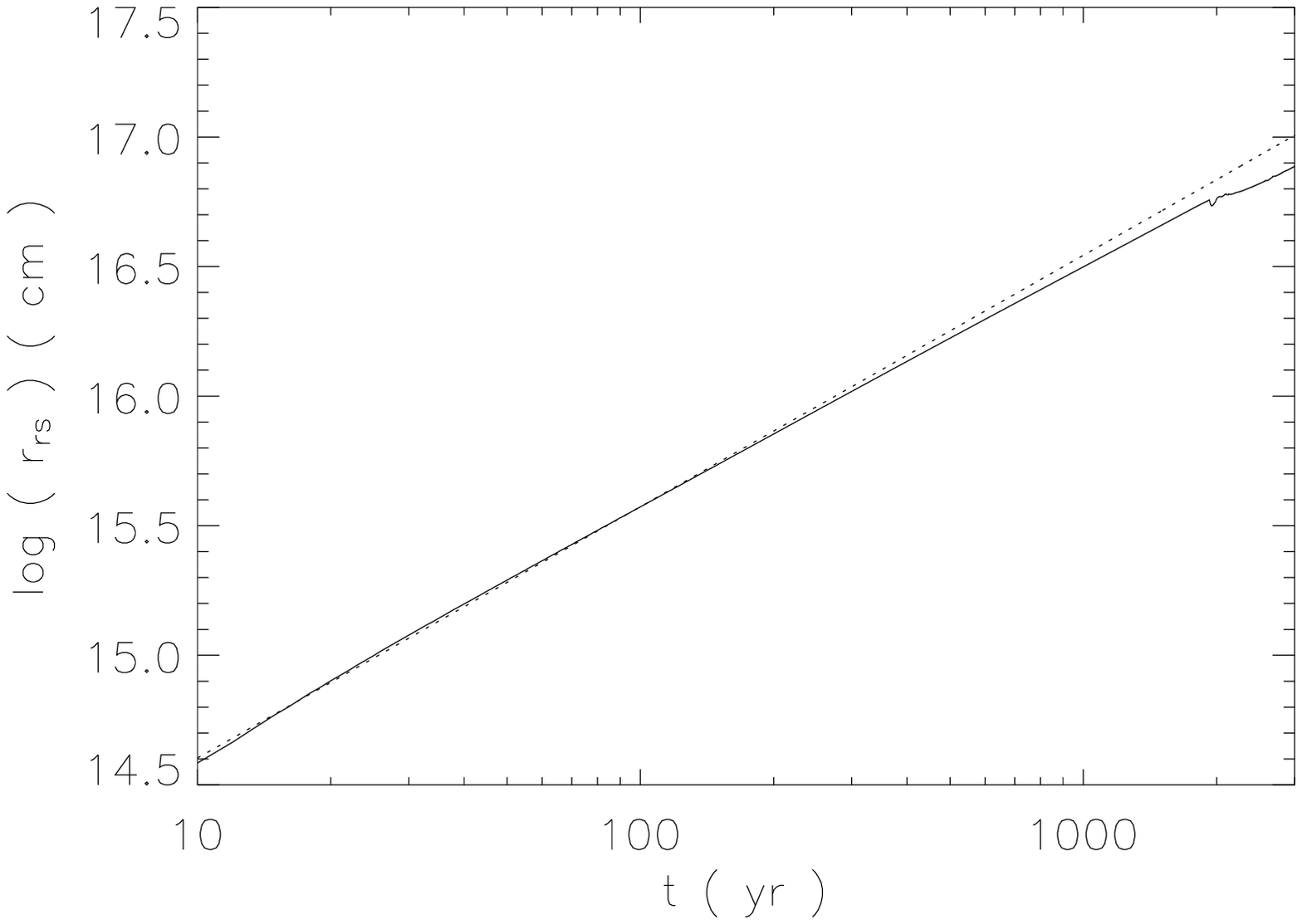}\\
\includegraphics[width=0.5\textwidth]{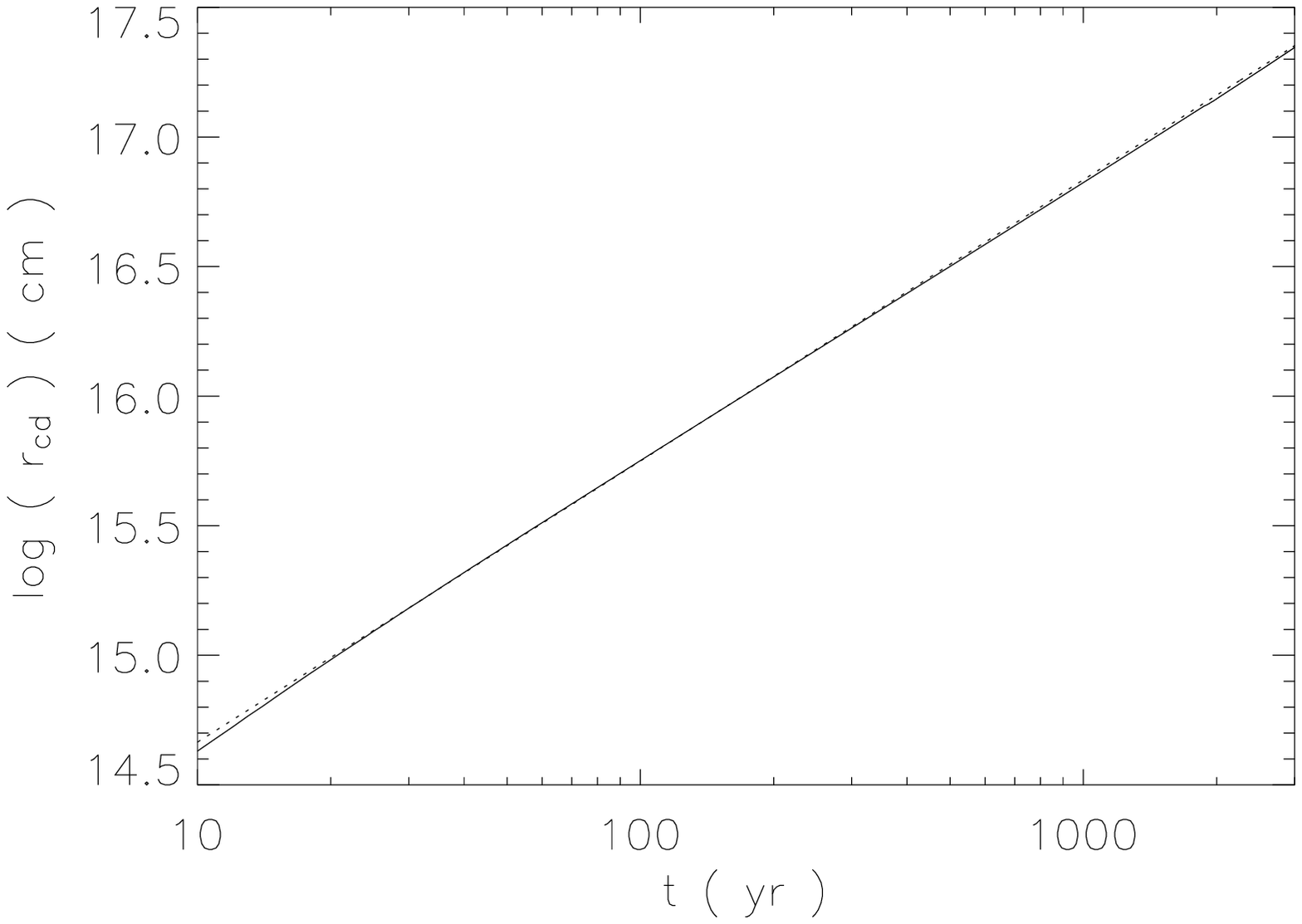}
\includegraphics[width=0.5\textwidth]{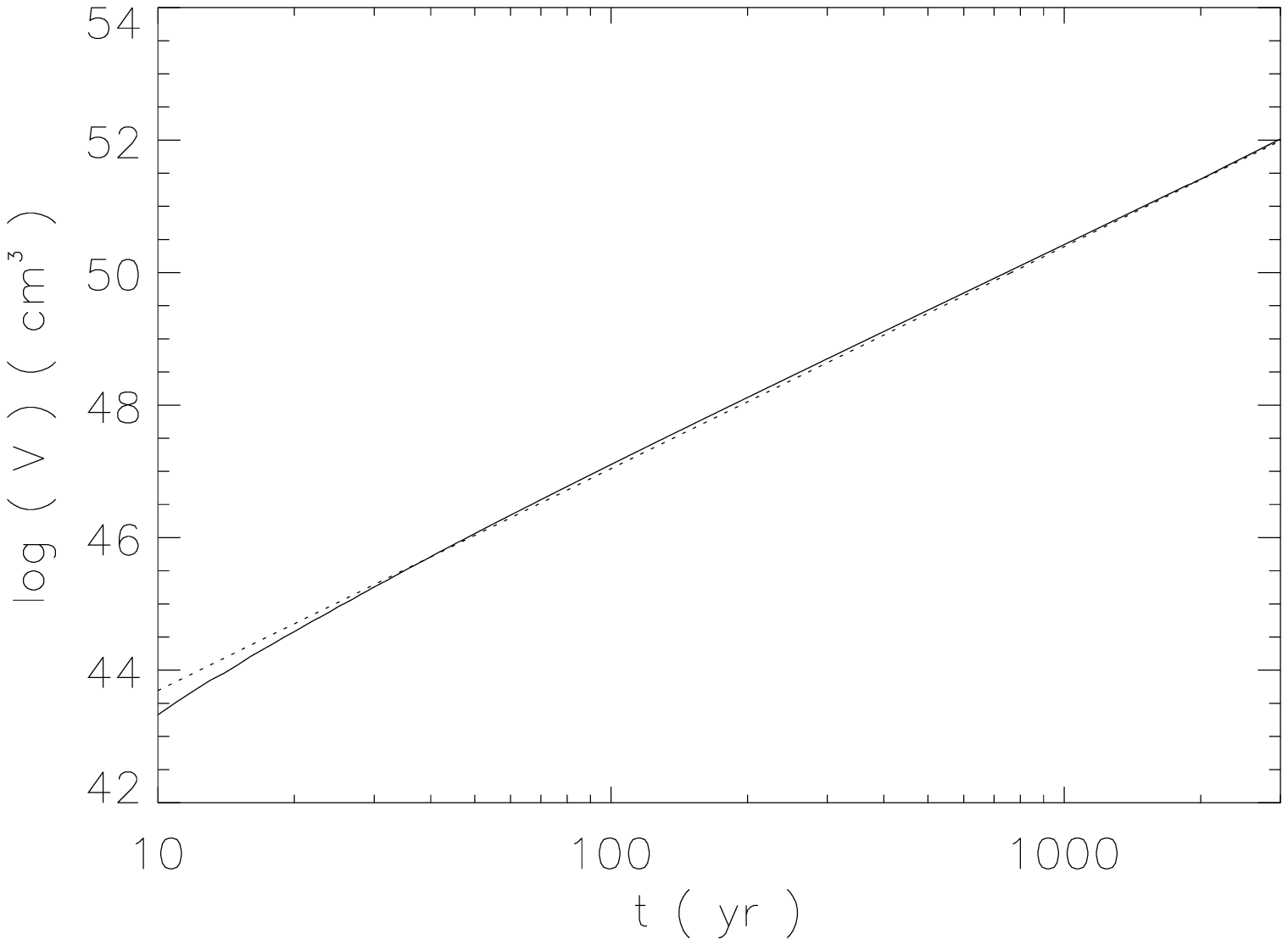}\\
\includegraphics[width=0.5\textwidth]{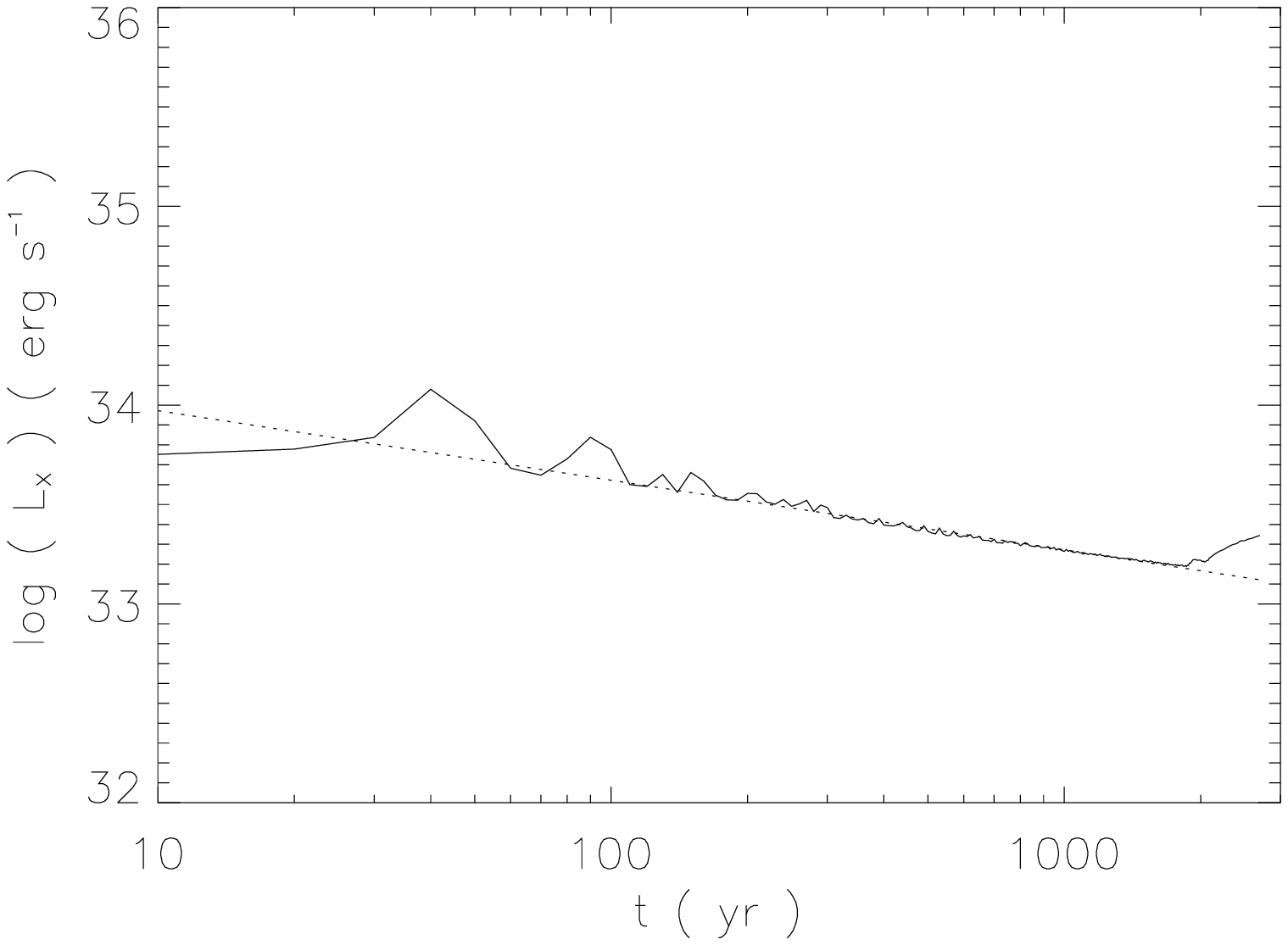}
\caption{Plots of the mean density $<\rho>$ of the hot bubble, the radius of 
the reverse shock $r_{\rm rs}$ and the contact discontinuity $r_{\rm cd}$
the volume of the hot bubble $V$ and of the X-ray luminosity $L_{\rm x,ACIS}$ 
in run E10 as a function of time; overplotted are the scaling laws for these 
variables ($< \rho > \sim t^{-1.86}$, $r_{\rm rs} \sim t^{0.98}$, $r_{\rm cd} 
\sim t^{1.08}$, $V = \left( r_{\rm cd}^3 - r_{\rm rs}^3 \right) \sim 
t^{3.35}$), which -- together with the temperature (not shown, as constant 
with time) -- determine the X-ray luminosity $L_{\rm x,ACIS}$} 
\label{fig_scalings}
\end{figure*}

We find that the higher the mass outflow rate of the fast post-AGB wind, the 
higher $L_{\rm x,ACIS}$ is (\S \ref{sec_res_1D_xray}), in accord 
with the results of ASB06. However, within each set of our runs with constant 
$\dot M_{\rm f}\,v_{\rm f}$, 
$L_{\rm x,ACIS}$ decreases for increasing $v_{\rm f}$ -- this behavior is 
different from the results of ASB06 in which the models with 
$v_{\rm f} = 500$ km s$^{-1}$ (A5, B5, C5) always have the highest 
$L_{\rm x,ACIS}$ value within each set of constant 
$\dot M_{\rm f}\,v_{\rm f}$. 

We can also compare the absolute luminosities in our models (which lie in the 
range $10^{30}$ -- $10^{33}$ erg s$^{-1}$) with those from 
ASB06\footnote{For this comparison, we have used bubbles with the same 
size rather than those with the same age, since the velocity of the contact 
discontinuity in our models is 
slightly different compared to ASB06. The velocities used by 
ASB06 follow from a cubic formula \citep[][also Table 1]{VoK85}.} (see
Table 2). We find good agreement (i.e. within a factor of 2--3) 
for the runs in sets A and B, 
however, in set C the derived $L_{\rm x,ACIS}$ values of ASB06 are higher 
(for example, a 
factor of 3 higher in run C5 and a factor of 5 higher in run C7). In 
addition, within each set (i.e. with constant 
$\dot M_{\rm f}\,v_{\rm f}$), in ASB06 the values of $L_{\rm x,ACIS}$ lie
within a factor $<2$ of each other, whereas in our models $L_{\rm x,ACIS}$ 
values cover a larger range of 4--10.

ASB06 conclude that their models can fit the X-ray properties and the 
dynamical ages of specific PNs, including BD$+$30$^\circ$3639 and NGC~40 
(e.g. Fig. 5 in ASB06). We think this conclusion is problematic for the
following reasons: first ASB06 use smaller values of the fast wind velocities 
than those which have been derived from measurements \citep{LHJ96}. They 
find that their best fits to BD$+$30$^\circ$3639 and NGC~40 are obtained with 
models C3 and B4, respectively, for which the fast 
wind velocities are too low compared with observations (300 and 400 
km s$^{-1}$ compared to 700 and 1000 km s$^{-1}$, respectively), 
and justify their choice by claiming that the details of the 
late-time evolution of the fast wind are not important to the X-ray emission.
They state that the X-ray emission predominantly emanates from wind 
segments of the hot bubble which were expelled early, i.e. with a moderate 
velocity, and over a relatively short time (few $\times$ 100 years). Hence 
the X-ray emission of the hot bubble is determined by the 
moderate fast wind and the subsequent time history is unimportant; however, we 
find that the reverse is true -- the X-ray emission is determined by the final 
stage of the time history during which the fast 
wind velocity has its largest value.

Second, their claimed agreement between models and data appears to break down,
when we consider in detail the model and observed values of all 
three parameters ($L_{\rm x}$, $T_{\rm x}$ and age) together. Thus,
although BD$+$30$^\circ$3639 is well fitted by C3 in the $\log ( 
L_{\rm x} )$--age plane (Fig. 5 of ASB06), in the 
$\log ( L_{\rm x} )$--$\log ( T_{\rm x} )$ plane (Fig. 6 of ASB06), 
it is far removed from model C3 and lies close to model B6.
NGC~40 (not shown in the $\log ( L_{\rm x} )$--age plane), is located
at a position in the $\log ( L_{\rm x} )$--$\log ( T_{\rm x} )$ plane which 
appears to lie on an extrapolation of the curve for model B4 representing
very late ages. However, taking into account the declining luminosity with 
increasing age and therefore size, this location would be reached at a size of 
the dense shell which is a factor 7 larger than observed.

\section{Fitting $T_{\rm x}$ and $L_{\rm x}$ in PNs -- an unsolved problem?}
\label{sec_fit}

In section \ref{sec_appl}, we have found that using the values of 
$v_{\rm exp}$, 
$r_{\rm shell}$, $\dot M_{\rm f}$ and $v_{\rm f}$ in  
BD$+$30$^\circ$3639 and NGC~40 as inferred from observations, our models 
produce values of $T_{\rm x}$ and 
$L_{\rm x, ACIS}$ which are much higher than the observed values. We explored 
three different mechanisms for reducing $T_{\rm x}$ and $L_{\rm x, ACIS}$. 
First, noting that the central stars in both objects are of [WC] type, 
deviations of the abundances from the solar values were tested as a way of 
reducing $L_{\rm x, ACIS}$. In the case of 
BD$+$30$^\circ$3639, this mechanism appears to be promising, but 
needs to be tested using a cooling function consistent with the non-solar 
abundances. In the case of NGC~40, although $L_{\rm x, ACIS}$ was reduced, 
the discrepancy between the model and the observed value remains
too large. The second mechanism, in which the fast wind had a slower velocity 
initially, was also not effective. Thirdly, heat conduction, which has been
discussed previously as a possible mechanism, does lower $T_{\rm x}$, but 
leads to a large increase in $L_{\rm x, ACIS}$ \citep{SSW06} due to an
increase in the density of the hot bubble. We also found 
that, even though $L_{\rm x, ACIS}$ is reduced in models computed on a 
two-dimensional grid, the reduction is not adequate. 

Since $T_{\rm x}$ ($L_{\rm x, ACIS}$) depends sensitively on $v_{\rm f}$ 
($\dot M_{\rm f}$), is it possible that the values of these parameters 
inferred from observations by \citet{LHJ96} are too high? Since $v_{\rm f}$ 
is directly measured from Doppler shifts of absorption features, the 
uncertainties in determining $v_{\rm f}$ are likely to be small. The value of 
$\dot M_{\rm f}$ is of course model-dependent and is also
affected by the assumed distance as $\dot M_{\rm f} \sim d^{3/2}$ 
\citep{LHJ96} -- because of the distance dependency, $L_{\rm x, ACIS}$ 
should vary with distance as $\sim d^{9/4}$, since in our 
models $L_{\rm x, ACIS} \sim \dot M_{\rm f}^{3/2}$ 
(Fig. \ref{Lvsmdotf}). $L_{\rm x, ACIS}$ in our model is also dependent on 
time and therefore on the size of the bubble. Assuming 
the dependency $L_{\rm x, ACIS} \sim t^{-0.37}$ (see \S \ref{sec_diff}), we get
$L_{\rm x, ACIS} \sim r^{-0.37} \sim d^{-0.37}$. Combining both dependencies, 
the ratio of $L_{\rm x, ACIS}$ in our models to the observed value varies as 
$\sim d^{2.25 - 0.37} / d^2 = d^{-0.12}$ -- because of this weak dependency, 
distance uncertainties cannot be invoked to reduce the discrepancies between 
the model and observed values of $L_{\rm x, ACIS}$. Using the dependency 
$L_{\rm x, ACIS} \sim \dot M_{\rm f}^{3/2}$ (Fig. \ref{Lvsmdotf}), 
$\dot M_{\rm f}$ would have to be reduced by a factor of 50 from the value
inferred by \citet{LHJ96} in order to achieve consistency between the model 
and the observed value of $L_{\rm x, ACIS}$ in NGC~40. Of course, the problem 
of the model $T_{\rm x}$ being too high would still remain. In summary, the 
problem of hydrodynamical models for the formation of PNs producing values of 
$T_{\rm x}$ and $L_{\rm x}$ which are too high compared to observed values 
remains unsolved.

\section{Conclusion} \label{sec_concl}

We computed the structure and evolution of, and X-ray emission from,  
numerical simulations of one-dimensional interacting winds models and compared 
the results with 
different analytical models. The kinematics and the evolution of our bubbles 
and the structure of the flows agrees well with analytical results.
Comparing the X-ray properties of the hot bubbles in our models with the
analytical results of ASB06, we find some agreements and many disagreements.
The disagreements which are both qualitative and quantitative in nature argue
for the necessity of using numerical simulations for understanding the 
X-ray properties of PNs. 

We further investigated the luminosities in the energy ranges other than the 
0.2 -- 10 keV range covered by {\em ACIS}. We find that most of the X-ray flux 
may emerge at energies less than 0.09 keV (wavelengths above 137 \AA), 
although we do not have reliable quantitative estimates of this flux from
our models. At these wavelengths, commonly labeled as extreme and far UV, 
observations of PNs with FUSE (900 -- 1200 \AA), EUVE (70 -- 760 \AA), 
IUE (1150 -- 3200 \AA) and GALEX for even longer wavelengths
(1350 -- 2800 \AA) would therefore be quite important.  
Unfortunately, photospheric 
emission from the hot central star easily blends with the emission from the PN.
Instruments with high spatial resolution are needed.

We also investigated the X-ray spectra from our models, which are not 
addressed by the analytical studies. We showed that the shapes of the spectra 
are diagnostic of the fast wind velocity which determines the temperature of 
the hot bubble.

We applied our spherical model to the objects BD$+$30$^\circ$3639 and NGC~40.
We find in both cases that the simulated temperatures and luminosities of the
hot bubble are much higher compared to the observations.
While investigating these discrepancies, we identified several general 
issues which have to be considered for correctly modeling the X-ray emission 
of PNs. 

The first one is the importance of an independent knowledge of the abundances 
-- especially in objects with [WC]-type central stars, where the abundances of 
the hot bubble (shocked fast wind) can be dramatically different from those in 
the dense shell (shocked slow wind). We could roughly reproduce the observed 
luminosity in 
BD$+$30$^\circ$3639 by simply (i.e. in a non-self-consistent manner) removing
several heavy elements including iron in the calculations of the X-ray 
properties. The treatment of cooling, however, has to be done 
self-consistently, as the cooling function itself is highly sensitive to 
abundances (e.g. iron). 

A second issue is the possible dependence of $L_{\rm x}$ and $T_{\rm x}$ 
on the time history of the fast wind velocity.
Noting that $T_{\rm x}$ is most sensitive to $v_{\rm f}$, we investigated 
whether $L_{\rm x}$ and $T_{\rm x}$ could be lowered by assuming that the 
value of $v_{\rm f}$ was lower in the past, but found that such a
time history will not help in explaining the observed low values of 
$L_{\rm x}$ and $T_{\rm x}$. The X-ray emission is determined by the final 
stage of the time history during which the fast 
wind velocity has its largest value. Although we have calculated only one case 
of wind evolution with a particular jump in the fast wind properties, the 
claim of ASB06, that the X-ray emission of the hot bubble is determined by the 
moderate fast wind and that the subsequent time history is unimportant, is not 
supported by our results.

We discussed the importance of heat conduction on lowering the temperature of 
the hot bubble, and its possible inhibition by magnetic fields. As the 
presence of magnetic fields is supported both by observations and theoretical 
considerations, heat conduction is likely inhibited. Even fields of the order 
of $\mu G$ are sufficient to inhibit heat conduction across (and along) the 
field lines. Since the X-ray spectra strongly indicate a very large 
overabundance of carbon in the hot bubble (as expected in the fast wind from 
the [WC] central star of this object, but much higher than the carbon 
abundance of the dense shell), if heat conduction is operating in this object,
it can not have evaporated a large amount of mass from the dense shell into 
the hot bubble. Alternatively, heat conduction is inhibited by magnetic fields 
in this object.

We also calculated a two-dimensional spherical model to examine the presence 
of instabilities and how they affect the structure of the hot bubble and the
dense shell and the resulting X-ray properties. Due to these instabilities, 
some fraction of the total energy is redirected into the kinetic energy of 
non-radial motion, which leads to slower expansion 
velocities of the shell (by about 20 \%) and a reduced luminosity between 
0.2 -- 10 keV (by $\sim 1.6$), although the temperature of the hot bubble 
does not change significantly.

From our detailed modeling, we conclude that the problem of hydrodynamical 
models for the formation of PNs producing values 
of $T_{\rm x}$ and $L_{\rm x}$ which are too high compared to observed values 
remains unsolved. The simulations in our study are the first step towards 
modeling PNs and PPNs detected in X-rays which show 
bipolar or asymmetric structures with varying degrees of collimation. Our 
preliminary results show that higher collimation leads to lower X-ray 
luminosities \citep{StS06}. Detailed results will be presented in a future 
publication. 

\acknowledgments
We thank V. Dwarkadas for help related to the use of the {\em FLASH} code and 
D. Sch\"onberner for fruitful discussions during the IAU 234 conference. We 
acknowledge the comments and suggestions by the referee, Noam Soker, which 
improved the manuscript. The software used in this work was in part developed 
by the DOE-supported ASC / Alliance Center for Astrophysical Thermonuclear 
Flashes at the University of Chicago. This work was partially funded by 
NASA/CHANDRA grants GO3-4019X and GO4-5163Z, and NASA/STScI grant 
HST-GO-10317.01-A. The research described in this publication was carried out 
at the Jet Propulsion Laboratory, California Institute of Technology, under a 
contract with the National Aeronautics and Space Administration.

\end{document}